\documentclass[12pt]{article}

\usepackage{graphicx}
\usepackage{amsmath}
\usepackage{amssymb}
\usepackage{multirow}
\usepackage{url}

\allowdisplaybreaks
\def\eq#1{(\ref{#1})}
\def\[#1\]{\begin{align}#1\end{align}}
\def\hQ{\hat Q}
\def\hP{\hat P}
\def\hH{\hat{\mathcal{H}}}
\def\hJ{\hat{\mathcal{J}}}

\def\ctJ{{\cal \tilde J}}
\def\cJ{{\cal J}}
\def\cH{{\cal H}}

\begin{document}

\begin{titlepage}
\title{
\hfill\parbox{4cm}{ \normalsize YITP-16-131}\\ 
\vspace{1cm} 
Mother canonical tensor model}
\author{
Gaurav Narain$^{a}$\footnote{gaunarain@itp.ac.cn}, 
Naoki Sasakura$^{b}$\footnote{sasakura@yukawa.kyoto-u.ac.jp}
\\
$^{a}${\small{\it Kavli Institute for Theoretical Physics China (KITPC)
\& Key Laboratory of Theoretical Physics, }}\\
{\small{\it
Institute of Theoretical Physics, 
Chinese Academy of Sciences (CAS), Beijing 100190, P.R. China.}}
\\
$^{b}${\small{\it Yukawa Institute for Theoretical Physics, Kyoto University,}}
\\ {\small{\it  Kitashirakawa, Sakyo-ku, Kyoto 606-8502, Japan,}}
}

\date{\today}
\maketitle
\thispagestyle{empty}
\begin{abstract}
\normalsize
Canonical tensor model (CTM) is a tensor model formulated in the Hamilton 
formalism as a totally constrained system with first class constraints, 
the algebraic structure of which is very similar to that of the ADM formalism of general relativity.  
It has recently been shown that a formal continuum limit of the classical equation of motion of CTM 
in a derivative expansion of the tensor up to the fourth derivatives agrees with that of a coupled system of 
general relativity and a scalar field in the Hamilton-Jacobi formalism. This suggests the existence of 
a ``mother" tensor model which derives CTM through the Hamilton-Jacobi procedure, and
we have successfully found such a ``mother" CTM (mCTM) in this paper.
The quantization of mCTM is straightforward as CTM.
However, we have not been able to identify all the secondary constraints, and therefore the 
full structure of the model has been left for future study.
Nonetheless, we have found some exact physical wave functions 
and classical phase spaces which can be shown to solve the primary and all the (possibly infinite) 
secondary constraints in the quantum and classical cases, respectively, and have thereby proven the non-triviality of the model.
It has also been shown that mCTM has more interesting dynamics than CTM 
from the perspective of randomly connected tensor networks.
\end{abstract}
\end{titlepage}

\section{Introduction}
Constructing quantum theory of gravity is one of the major fundamental problems in physics. 
Although we do not have such a fundamental theory yet, 
a number of thought experiments have been considered 
to grasp the essential picture of quantum gravity \cite{Garay:1994en} 
by qualitatively thinking of quantum gravitational effects. 
A common implication of these thought experiments seems
that the classical picture of smooth and continuum spacetime 
may not be valid in quantum gravitational regime. 
This would indicate that spacetime should be described by a new quantum notion
for successful construction of quantum gravity. 
In fact, a number of authors have proposed various models of quantum gravity
based on discretized building blocks of spacetime in the Planck scale. 
In these models, a spacetime in the classical picture 
should emerge as an infrared collective phenomenon of the dynamics of such building blocks.
Whether this is achieved or not gives an objective criterion for screening models.

The simplicial quantum gravity is one of such discretized approaches.
In this approach, a spacetime is modeled by gluing simplices,
and the random sum over them gives its quantization.
The tensor models were originally introduced as analytic
description of the simplicial quantum gravity in dimension 
$d>2$ \cite{Ambjorn:1990ge,Sasakura:1990fs,Godfrey:1990dt}\footnote{However, 
see \cite{Fukuma:2015xja,Fukuma:2015haa,Fukuma:2016zea} 
for a matrix-model-like approach to the $d=3$ simplicial quantum gravity.}, 
hoping to extend the success of the matrix models in $d=2$.
While these original tensor models are still remaining merely as sort of formal description, 
colored tensor models \cite{Gurau:2009tw}
have produced a number of interesting analytical results
in the large $N$ limit \cite{Gurau:2011xp}. 
Among them, it has been shown that the simplicial spaces 
generated from the colored tensor models are dominated by 
branched polymers \cite{Bonzom:2011zz,Gurau:2013cbh}.
Since branched polymers have very different structures from our actual spacetime, 
further improvement would be necessary for tensor models to be qualified as quantum 
gravity\footnote{However, tensor models  
are recently attracting some attentions 
as SYK-like models without disorder \cite{Witten:2016iux,Klebanov:2016xxf} in the context of AdS/CFT correspondence.
In this context, the dominance of such diagrams plays essential roles, and  
tensor models may be related with quantum gravity through duality.}.

Similar results have been obtained from the numerical analysis of the 
simplicial quantum gravity. In Dynamical Triangulation (DT),
the random sum over simplicial spaces is not dominated by the ones 
consistent with the classical picture of spacetime. 
On the other hand, the model with a causality, called Causal Dynamical Triangulation (CDT),
has been shown to produce macroscopic spacetimes, 
which are similar to the de Sitter spacetime like our universe \cite{Ambjorn:2004qm}. 
The essential difference between DT and CDT is that, while the former 
basically concerns Euclidean simplicial spaces, the latter
incorporates causality to generate Lorentzian simplicial spaces. 
Here, the condition of causality prohibits the topology change of 
spatial slices of Lorentzian simplicial spaces,
which is the source of the difficulty in the Euclidean case.  
 
The above success of CDT over DT suggests that causality would essentially be 
important for the emergence of classical spaces in formulating quantum gravity.
Causality can naturally be incorporated by using Hamilton formalism. 
This motivated one of the present authors to construct a tensor model in Hamilton 
formalism \cite{Sasakura:2011sq,Sasakura:2012fb}.
A concern about the use of Hamilton formalism for this purpose is that, 
if a time direction was introduced in an explicit manner\footnote{This is actually taken in the 
SYK-like models \cite{Witten:2016iux,Klebanov:2016xxf}.},
the dynamics of an emergent spacetime, if it appeared, would not be described in a spacetime covariant manner,
contradicting the central principle of general relativity. 
Therefore, as in the ADM formalism of general relativity 
\cite{Arnowitt:1960es,Arnowitt:1962hi,Hojman:1976vp}, a time direction
must be introduced as a sort of gauge direction: the Hamiltonian of 
the model must be purely given by a linear combination
of first-class constraints. Remarkably, a tensor model can be 
formulated in this fashion \cite{Sasakura:2011sq}, 
which we call canonical tensor model (CTM), 
and moreover, under some physically reasonable assumptions, 
the model can be shown to be unique \cite{Sasakura:2012fb}. 

In fact, CTM has various intriguing properties. 
As for its relation to general relativity, the following properties have been shown.
(i) In the simplest case\footnote{This is the case of $N=1$, 
where $N$ denotes the range of tensor
indices taking $1,2,\cdots,N$.}, CTM classically agrees with 
the mini-superspace approximation of general 
relativity \cite{Sasakura:2014gia}. (ii) In a formal continuum limit with $N\rightarrow \infty$, 
the algebraic structure of the constraints of CTM agrees with 
that of the ADM formalism \cite{Sasakura:2015pxa}.
(iii) (ii) can be detailed further.
In a derivative expansion up to the fourth order in the formal 
continuum limit, the classical equation of motion of CTM 
agrees with that of a coupled system of general relativity 
and a scalar field in the Hamilton-Jacobi formalism \cite{Chen:2016ate}. 
Presently, we do not know whether CTM can successfully 
generate classical spaces as infrared phenomena of its dynamics, 
but the above results seem to imply that, if a classical space emerged from CTM, 
the dynamics in such a space 
would be described by a general relativistic system. 

Another sort of intriguing results come from its connection to the randomly connected tensor networks 
(RCTN) \cite{Sasakura:2014zwa,Sasakura:2014yoa,Sasakura:2015xxa,revnetwork}. 
RCTN is a tensor network with random connections, and 
can describe statistical systems on random networks \cite{revnetwork} 
such as Ising/Potts models on random networks, etc., by tuning tensors \cite{Sasakura:2014zwa,Sasakura:2014yoa}.
In the infinite size limit of networks, namely, in the thermodynamic limit, 
RCTN shows critical phenomena, and it would be interesting to 
seek for a renormalization group procedure for RCTN. 
Interestingly, while it is not clear how to define coarse-graining 
procedures on such dynamical networks \cite{revnetwork}, 
one can find that the Hamiltonian of CTM defines flows, which qualitatively agree with 
what renormalization group flows are supposed to be \cite{Sasakura:2014zwa,Sasakura:2015xxa}.    
An insight from RCTN was also very useful in finding the exact physical wave functions
of the quantized CTM \cite{Narain:2014cya}.
We would also like to comment that networks have appeared in some other approaches to 
spacetime \cite{Penrose:1971}-\cite{Bianconi:2015hfa}.
In this respect, CTM describes spacetime in terms of networks rather than 
simplicial spaces, differently from the original motivation of the tensor models.

The quantization of CTM is straightforward \cite{Sasakura:2013wza}. 
The constraints of CTM remain first-class, even after quantization.
An important physical question in the quantized case is to obtain the 
physical wave functions which are defined as 
the common kernel of the quantized constraints. 
The condition can be represented by a set of partial differential equations for a wave function.
The first-class nature, namely, the algebraic closure of the constraints, 
ensures the existence of solutions, but there are no apparent 
reasons for analytical solvability. Nonetheless, quite remarkably, we have 
obtained various exact solutions with analytic 
expressions \cite{Narain:2014cya,Sasakura:2013wza}.
As mentioned above, an insight from RCTN was quite useful in 
finding some general series of exact solutions \cite{Narain:2014cya}.  

\begin{figure}
\begin{center}
\includegraphics[scale=1]{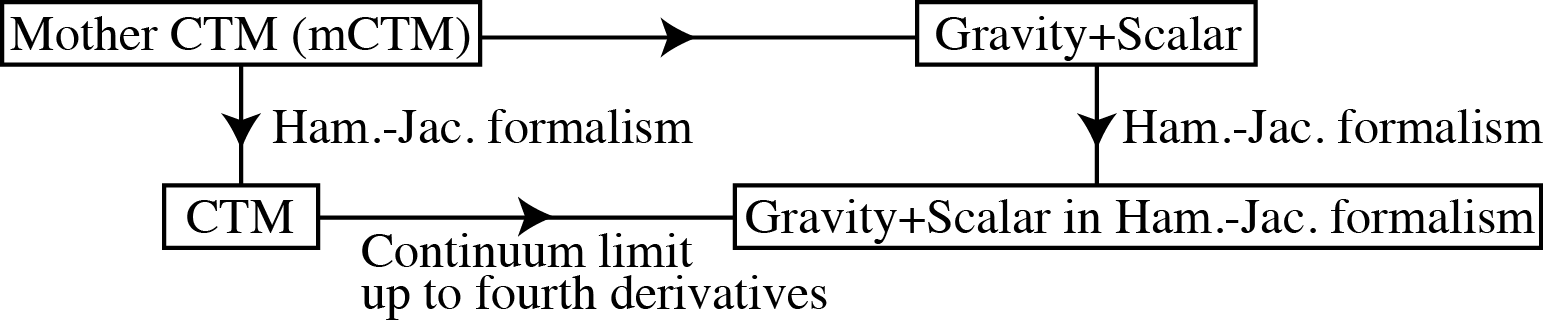}
\caption{A schematic representation of the motivation}
\label{fig:scheme}
\end{center}
\end{figure}
The main purpose of this paper is to present a new formulation of CTM. 
This is motivated from the aforementioned identification 
of a formal continuum limit of the classical CTM with a general relativistic system 
in the Hamilton-Jacobi formalism  \cite{Chen:2016ate}. 
There, the identification has been done with the most natural but a particular form of
the Hamilton's principal function of a general relativistic system. This means that
the classical CTM can be identified with part of the dynamics of the general relativistic system,
but not all of it.
Therefore, we want to obtain a tensor model which would be more directly 
related to a general relativistic system without resorting to the 
Hamilton-Jacobi formalism (See Figure~\ref{fig:scheme}). 
In this paper, we will show that there actually exists a ``mother" CTM (mCTM) which gives CTM 
through the Hamilton-Jacobi procedure.

We will also show the results of our initial study on the properties of mCTM, 
while its continuum limit is left for future study.
As we will see, there exists a crucial difference between mCTM and CTM. 
While CTM has a closed algebra of two kinds of constraints, 
the closure of the constraints is not obvious in mCTM:
it is not clear what are the independent set of the secondary constraints 
and it may even be possible that there are an infinite number of them.
In such a situation, the existence of a nontrivial dynamical solution consistent 
with all the constraints is not guaranteed, as the restrictions imposed 
by constraints might be too tight.
In this paper, we cannot answer this question in general terms,
but will show some concrete evidences for the non-triviality of the dynamics of mCTM:  
we will obtain some exact physical wave functions in the quantum case, and 
also obtain some non-trivial phase-space 
solutions in the classical case,
which solve the primary and all the (possibly an infinite number of) secondary constraints. 

This paper is organized as follows. 
In the following section, we give a brief review of CTM. 
In Section~\ref{sec:new}, we present mCTM which leads to CTM through the Hamilton-Jacobi procedure.
In Section~\ref{sec:quantum} and Section~\ref{sec:classical}, 
we show some non-trivial solutions
to all the constraints in the quantum and classical cases, respectively.
In Section~\ref{sec:network}, we  show 
that mCTM is a natural extension of CTM from the perspective of tensor network dynamics.
Section~\ref{sec:summary} is devoted to a summary and an outlook.

\section{Review of CTM}
\label{sec:review}
In this section, we will review the canonical tensor model (CTM) 
\cite{Sasakura:2011sq, Sasakura:2012fb}
to make this paper self-contained and to fix the notations to be used.

CTM is a Hamiltonian system with a canonical conjugate pair of rank-three 
tensors\footnote{In this paper, we use the more familiar notation 
$Q$ and $P$ as the conjugate pair of variables, 
instead of $M$ and $P$, which has been used in the previous papers. The familiar notation would be more appropriate, 
because the conjugate pair appears symmetrically in the new formulation.},
$Q_{abc}$ and $P_{abc}$ $(a,b,c=1,2,\cdots, N)$, which are assumed to be 
real and symmetric under index permuations. 
The fundamental Poisson brackets are given by
\[
\{ 
Q_{abc}, P_{def}
\}
= 
\sum_{\sigma}
\delta_{a \sigma_d}\delta_{b \sigma_e}\delta_{c \sigma_f}, 
\ \ \ 
\{ 
Q_{abc}, Q_{def}
\}
= 
\{ 
P_{abc}, P_{def}
\}
=0,  
\label{eq:poisson}
\]     
where the summation is over all the permutations of $d$, $e$ and $f$, 
incorporating the symmetric nature of the tensors. 
The kinematical symmetry of CTM is assumed to be given by the 
invariance under the $O(N)$ symmetry 
defined by  
\[
\begin{split}
&Q_{abc} \to Q'_{abc} = L_{aa'}L_{bb'}L_{cc'}Q_{a'b'c'}, \\
&P_{abc} \to P'_{abc} = L_{aa'}L_{bb'}L_{cc'}P_{a'b'c'},
\end{split}
\label{eq:ontransformation}
\]
where repeated indices are summed over and $L$ denotes
the $O(N)$ matrices. The transformation \eq{eq:ontransformation} 
is consistent with the aforementioned properties of $Q$ and $P$,
and pair-wise contractions of their indices are invariant under the transformation.

As briefly explained in introduction, it is necessary to incorporate 
time as a gauge direction due to the restriction imposed by 
general covariance of the emergent spacetime.   
This is indeed realized in the Hamilton formalism of general relativity, 
so called Arnowitt-Deser-Misner (ADM) formalism \cite{Arnowitt:1960es,Arnowitt:1962hi}, 
in which the Hamiltonian is purely given by a linear combination of first-class constraints. 
Similarly in CTM,  the Hamiltonian is given by
\[
H_{CTM}
= n_a \mathcal{H}_a + n_{ab} \mathcal{J}_{ab},
\label{eq:ctmhamiltonian}
\] 
where $n_a$ and $n_{ab}\ (=-n_{ba})$ are non-dynamical Lagrange's multipliers,
and $\mathcal{H}_a$ and $\mathcal{J}_{ab}$ are the first-class constraints of CTM.
We call $\mathcal{H}$ and $\mathcal{J}$ Hamiltonian and momentum constraints, respectively,
following the names used in the ADM formalism. 
We may also call $n_a$ a lapse (vector).
The constraints of CTM are given by
\[
\begin{split}
&\mathcal{H}_a 
=\frac{1}{2} \left(P_{abc}P_{bde}Q_{cde} - \lambda Q_{abb} \right), 
\\
&\mathcal{J}_{ab}
=-\mathcal{J}_{ba}
= \frac{1}{4} \left( P_{acd}Q_{bcd} - P_{bcd}Q_{acd} \right),
\end{split} 
\label{eq:ctmconstraints} 
\]
where $\lambda$ is a real constant\footnote{In \cite{Sasakura:2014gia}, 
this parameter plays the role of the cosmological constant in the 
mini-superspace approximation of general relativity.}. 
By rescaling $Q$ and $P$ consistently with \eq{eq:poisson},
one can arrange to have $\lambda=0,\pm 1$ without loss of generality.
The classical equation of motion of CTM is given by
\[
\begin{split}
&\frac{d}{dt} Q_{abc} = \{Q_{abc}, H_{CTM} \} =\frac{1}{2} \sum_{\sigma}
\left(n_{\sigma_{a}} P_{\sigma_{b}de} Q_{\sigma_{c}de} 
+ n_d P_{de \sigma_{a}} Q_{\sigma_{b}\sigma_{c}e}
+  n_{\sigma_{a}d} Q_{\sigma_{b}\sigma_c d}\right),  \\
&\frac{d}{dt} P_{abc} 
= \{P_{abc}, H_{CTM}\} 
=  -\frac{1}{2} 
\sum_{\sigma}
\left(
n_d P_{de \sigma_{a}} P_{\sigma_{b}\sigma_{c}e} 
- \lambda n_{\sigma_{a}} \delta_{\sigma_{b}\sigma_{c}} 
+  n_{d \sigma_{a} } P_{\sigma_{b}\sigma_{c}d}
\right), 
\label{eq:ctmeom}  
\end{split}
\]
where $t$ is a time variable.

These constraints satisfy the following first-class Poisson algebra: 
\[
\begin{split}
& \{ \mathcal{H} (\xi^1), \mathcal{H} (\xi^2) \} 
= \mathcal{J} \left( [ \tilde{\xi}^1, \tilde{\xi}^2 ] + 2 \lambda\,  \xi^1 \wedge \xi^2 \right), \\
& \{ \mathcal{J} (\eta), \mathcal{H} (\xi) \} 
= \mathcal{H} (\eta \xi), \\
&\{ \mathcal{J} (\eta^1), \mathcal{J} (\eta^2) \} 
= \mathcal{J} \left( [ \eta^1, \eta^2 ] \right), 
\label{eq:ctmconstraintalgebra}
\end{split}
\]
where $\mathcal{H}(\xi) = \xi_a \mathcal{H}_a$, $\mathcal{J}(\eta) = \eta_{ab}\mathcal{J}_{ab}$, 
and $\tilde{\xi}_{ab}= P_{abc}\xi_c$. 
In (\ref{eq:ctmconstraintalgebra}), the bracket $[\ , \ ]$ denotes the matrix commutator, 
and $(\xi^1\wedge \xi^2)_{ab}= \xi^1_a \xi^2_b - \xi^2_a \xi^1_b$. 
$\mathcal{J}$ serves as the generators of $SO(N)$,
infinitesimally representing the kinematical symmetry \eq{eq:ontransformation}.
An important property of this algebra is that  
there exists non-linearity on the righthand side of the first equation in \eq{eq:ctmconstraintalgebra}
due to the $P$-dependence of $\tilde \xi$.
This property is in parallel with that of the algebra in the ADM formalism, where 
the Poisson bracket of the Hamiltonian constraints is dependent 
on the inverse metric tensor\footnote{See for example \cite{Hojman:1976vp} 
for comprehensive discussions on this aspect of general relativity.}.   

\begin{figure}
\begin{center}
\includegraphics[scale=1]{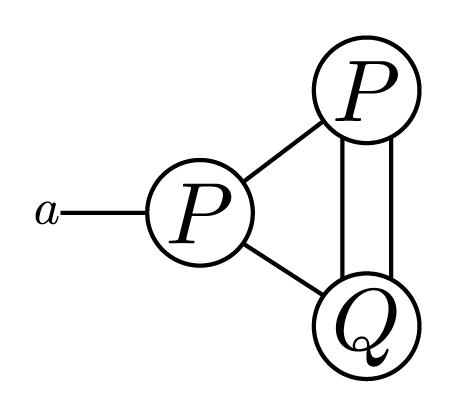}
\hfil
\includegraphics[scale=1]{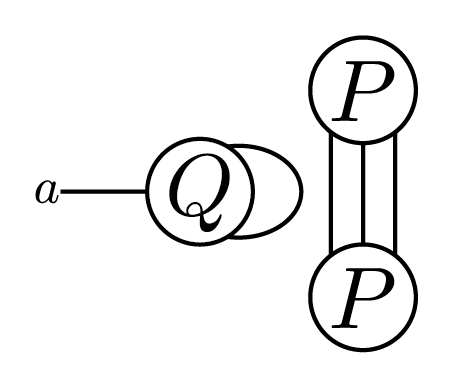}
\caption{The left and right figures show the examples of a connected and a disconnected term, respectively.}
\label{fig:connected}
\end{center}
\end{figure}

In fact, the model, and hence the constraint algebra as well, are 
unique under the following assumptions \cite{Sasakura:2012fb}:
(i) The dynamical variables are a conjugate pair of real symmetric rank-three tensors, $Q_{abc},P_{abc}$. 
(ii) There is a constraint which has one index, $\mathcal{H}_a$. 
(iii) There is a kinematical symmetry generated by the $SO(N)$ generators $\mathcal{J}_{ab}$. 
(iv)  $\mathcal{H}$ and $\mathcal{J}$ form a first-class constraint Poisson algebra.
(v) $\mathcal{H}$ is invariant under the time reversal 
transformation, $Q\rightarrow Q,\ P\rightarrow -P$. 
(vi) $\mathcal{H}$ is at most cubic in the canonical variables.
(vii) A tensor-model analogue of locality is respected: there should be 
no disconnected terms in the constraint algebra: e.g., 
$P_{abc}P_{bde}Q_{cde}$ is allowed but $Q_{abb}P_{cde}P_{cde}$ is not (See Figure~\ref{fig:connected}.).

Among all the assumptions above, the standpoint of (vi) seems fragile, while the others are more or less physically 
motivated \cite{Sasakura:2011sq,Sasakura:2012fb}. It was assumed in \cite{Sasakura:2012fb} simply for
technical limitations in showing the closure of the constraint algebra, 
because the computations of the Poisson brackets among all the possible forms of constraints
became quite cumbersome even under the limitation (vi). To remove (or relax) (vi), 
it would be necessary to develop a more systematic methodology for analysis than the brute force one which has been
done in \cite{Sasakura:2012fb}. Therefore, at present, we cannot prove or disprove whether the 
tensor model is unique 
even after the removal (or relaxation) of (vi) in our line of gauging the time direction in 
Hamilton formalism.  However, we would be able to speculate the following matters in general.
Closure of constraint algebras would become more difficult to be realized for higher order 
interactions or higher rank tensor models, 
because Poisson brackets among constraints would generate more numerous and more complex terms, 
which, for algebraic closure, must end up to be proportional to constraints after numerous cancellations. 
This suggests that, even if other tensor models with a gauged time direction
existed for higher order interactions or ranks,
the possibilities would strongly be limited. This aspect is largely different from the other tensor models which
do not gauge the time direction, because, in these models, one can add various higher order terms and 
higher rank tensors with basically free choices under rather weak requirements for kinematical symmetries.    
In addition, it is wroth stressing that, differently from the other tensor models, 
the rank-three of the tensors of CTM does not restrict the dimension of spacetimes to be three.
This is due to the distinction of the spacetime interpretation of CTM from the others, and 
has actually been explicitly shown in the results so far summarized below.  In this sense, 
the rank-three is enough for physical purposes.

The studies so far have shown some remarkable connections 
between the classical framework of CTM and general relativity in arbitrary dimensions. 
Firstly, for $N=1$, the Hamiltonian (\ref{eq:ctmhamiltonian}) agrees 
with that of a certain mini-superspace model of general relativity in arbitrary dimensions, 
if we consider the modulus of the tensor, $|Q_{111}|$, is proportional to the 
spatial volume in the mini-superspace model \cite{Sasakura:2014gia}. 
Secondly, in a formal continuum limit with $N \to \infty$, 
the Poisson algebra (\ref{eq:ctmconstraintalgebra}) coincides with 
the algebra of constraints in the ADM formalism \cite{Sasakura:2015pxa}.
In fact, this coincidence can be studied further, and it has been shown,
through a derivative expansion of $P$ up to the fourth order, that the 
classical equation of motion of $P$ in (\ref{eq:ctmeom}) in the formal continuum limit
can be identified with that of a coupled system of general relativity and a scalar field
in the Hamilton-Jacobi formalism \cite{Chen:2016ate}.
The successes above concerning the connection between the classical CTM and general relativity 
gave us a good motivation to further investigate the model 
quantum mechanically as a model for quantum gravity.  

The quantization of CTM is straightforward \cite{Sasakura:2013wza}. Let us promote 
\eq{eq:poisson} to the following fundamental commutation relations:
\[
[ \hQ_{abc}, \hP_{def} ] 
=i \sum_{\sigma}
 \delta_{a\sigma_d} \delta_{b\sigma_e} \delta_{c\sigma_f},
 \ \ 
 [ \hQ_{abc}, \hQ_{def} ]=[ \hP_{abc}, \hP_{def} ]=0.
 \label{eq:commutator}
\] 
Then, the quantized Hamiltonian is given by
\[
\hat{H}=n_a \hat{\mathcal{H}}_a +n_{ab} \hat{\mathcal{J}}_{ab}, 
\label{eq:qH}
\] 
where 
\[
\begin{split}
&\hat{\mathcal{H}}_a = \frac{1}{2} \left( \hat{P}_{abc} \hat{P}_{bde} \hat{Q}_{cde} 
- \lambda \hat{Q}_{abb} + i \lambda_{H} \hat{P}_{abb}   \right),  \\
&\hat{\mathcal{J}}_{ab} =-\hat{\mathcal{J}}_{ba} = \frac{1}{4} \left( \hat{P}_{acd} \hat{Q}_{bcd} 
- \hat{P}_{bcd} \hat{Q}_{acd}  \right).
\end{split}
\label{eq:qhj}
\]
There is a new term with a real constant $\lambda_H$, which comes 
from the normal-ordering of the first term of $\hH$. 
By demanding the hermiticity of $\hat{\mathcal{H}}$, the proportional constant 
is determined to be $\lambda_H = (N+2)(N+3)/2$.

A non-obvious convenient fact about the quantization is that the 
constraint algebra does not change from the classical case.
Namely, the constraints remain first-class, and the consistency of quantum CTM is guaranteed.
In fact, from explicit computations, one obtains
\[
\begin{split}
&[\hH (\xi^1), \hH (\xi^2)] = i\,\hJ \left( [\hat{\xi^1}, \hat{\xi^2} ] + 2 \lambda\, \xi^1 \wedge \xi^2  \right),\\
&[\hJ (\eta), \hH (\xi)] = i\,\hH \left(\eta \xi \right), \\
&[\hJ (\eta^1), \hJ (\eta^2)] = i\, \hJ \left( [\eta^1,\eta^2] \right)
\end{split}
\label{eq:qalg}
\]
where $\hH(\xi)= \xi_a\hH_a$, $\hJ (\eta)= \eta_{ab}\hJ_{ab}$, and 
$\hat{\xi}_{ab}= \hP_{abc}\xi_c$ with $c$-numbers $\xi_a,\eta_{ab}$. 
Here, the ordering is assumed to be $\hJ (\hat \eta):= \hat \eta_{ab}\hJ_{ab}$, 
when $\hat \eta_{ab}$ is an operator, as on the righthand side of the first line.
   
A physical state $\Psi$ of quantum CTM is defined by 
\[
\hH_a  \Psi=\hJ_{ab} \Psi =0.
\label{eq:physstates}
\]
Because of the closure of the algebra \eq{eq:qalg}, no additional 
secondary constraints will be imposed on the physical states.
Since the number of the constraints is not larger than the dimension of the configuration space, 
there exist such physical states in general.
By choosing a representation, \eq{eq:physstates} can be expressed 
as a set of partial differential equations, which generally have very 
complicated forms. Nonetheless, various exact solutions have been 
found \cite{Sasakura:2013wza,Narain:2014cya}.
Presently, this analytical solvability seems mysterious.

\section{Mother  CTM}
\label{sec:new}
In this section, we will obtain a ``mother" CTM (mCTM) which 
derives CTM through the Hamilton-Jacobi procedure. 

Let us first recall the Hamilton-Jacobi formalism. Let us consider a classical system with 
dynamical variables $x_i(t)\ (i=1,2,\cdots,n)$, where $t$ is a time variable.
Then, let us consider an action,
\[
S=\int dt \ \left( 
\frac{1}{2} K_{ij}(x) \left(\dot x_i-L_i(x)\right) \left(\dot x_j-L_j(x)\right)-V(x)
\right),
\label{eq:classicalaction}
\]
where $K,L$ are assumed to be independent of $t$.
The conjugate momenta of $x$ are given by
\[
p_i=K_{ij}(x)\left(\dot x_j-L_j(x)\right).
\label{eq:classicaldefp}
\]
Then, the Hamiltonian is given by
\[
H=\frac{1}{2}  K(x)^{-1}_{ij} p_i p_j+p_i L_i(x) +V(x).
\label{eq:classicalhamiltonian}
\]

The Hamilton-Jacobi formalism replaces the problem of solving the equation of motion with
solving
a partial differential equation called Hamilton-Jacobi equation. This is obtained by considering 
a canonical transformation with
\[
p_i=\frac{\partial W(x)}{\partial x_i},
\label{eq:classicalpW}
\]
and putting \eq{eq:classicalpW} into the Hamiltonian \eq{eq:classicalhamiltonian}:
\[
\frac{1}{2} K(x)^{-1}_{ij} \frac{\partial W(x)}{\partial x_i} \frac{\partial W(x)}{\partial x_j}
+\frac{\partial W(x)}{\partial x_i}L_i(x)+V(x)=E,
\label{eq:HJequation}
\]
where $W(x)$ is called Hamilton's principal function and $E$ is a constant.
When $W(x)$ is obtained as a function of $x$ by solving \eq{eq:HJequation}, 
the trajectory of $x$ is determined from
\[
\dot x_i =K(x)^{-1}_{ij} \frac{\partial W(x)}{\partial x_j}+L_i(x),
\label{eq:eqofx}
\]
which has been obtained from \eq{eq:classicaldefp} and \eq{eq:classicalpW}.
This is a first-order differential equation in time of $x$,
while the equation of motion of $x$ derived from \eq{eq:classicalaction} is second-order if 
it is expressed solely in $x$.

Our main idea in this paper is to identify \eq{eq:eqofx} with the classical equation of motion of $P$ in \eq{eq:ctmeom},
which is a first-order differential equation in time. 
A similar idea was used in the previous paper \cite{Chen:2016ate} 
to show the agreement between CTM in a continuum limit and a gravitational system
in the Hamilton-Jacobi formalism. The difference between the paper \cite{Chen:2016ate}  and 
the present one is that we will apply the idea directly to CTM without considering a continuum limit nor general relativity.
Namely, we want to find a ``mother" CTM (mCTM) which gives CTM through the Hamilton-Jacobi 
procedure (See Figure~\ref{fig:scheme}.).
  
In the present case, $P_{abc}$ correspond to $x_i$ with $i$ corresponding to $abc$.
Then, one can easily find that
\[
\begin{split}
K^{-1}_{abc,def}&=\frac{1}{6} \sum_{\sigma,\sigma'} n_g P_{g\sigma_a \sigma'_d} \delta_{\sigma_b\sigma'_e} 
\delta_{\sigma_c \sigma'_f},\\
W&=\frac{1}{2} P_{abc}P_{abc}, \\
L_{abc}&=\sum_{\sigma}\left( \lambda n_{\sigma_a} \delta_{\sigma_b\sigma_c}+n_{\sigma_a d}P_{d \sigma_b \sigma_c}
\right),
\end{split}
\label{eq:KandW}
\]
reproduces the equation of motion of $P$ in \eq{eq:ctmeom},
where we have included unimportant numerical coefficients into $n_a,n_{ab},\lambda$ 
for simplicity of the expression.

From \eq{eq:HJequation} and \eq{eq:KandW}, one can determine the potential $V$ as
\[
V= - 3 n_a P_{abc}P_{bde}P_{cde} -6 \lambda n_a P_{abb},
\label{eq:V}
\]
where we have set $E=0$, because such a constant cannot be expressed in an invariant manner  
proportional to $n_a$ or $n_{ab}$. 
Here, the second term of $L$ in \eq{eq:KandW} does not contribute to $V$, because of the antisymmetry, $n_{ab}=-n_{ba}$
(See Section~\ref{sec:review}). 
By putting $p_i=-Q_{abc}$, \eq{eq:KandW}, and \eq{eq:V} into \eq{eq:classicalhamiltonian}, one obtains
\[
H=3 n_a P_{abc}Q_{bde}Q_{cde} -6 \lambda n_a Q_{abb}-6 n_{ab} Q_{acd}P_{bcd}
- 3 n_a P_{abc}P_{bde}P_{cde} -6 \lambda n_a P_{abb}.
\label{eq:HofCTM}
\]
Note that, because of the antisymmetry of $n_{ab}$, the third term of \eq{eq:HofCTM} 
is nothing but the $SO(N)$ generator
$n_{ab}{\cal J}_{ab}$, and therefore can be separated from the ``Hamiltonian" constraint.  

The expression \eq{eq:HofCTM} looks unusual due to the cubic term of $P$, but can be made
quadratic in $P$, if we consider a canonical transformation, $Q\rightarrow - c P,\ P\rightarrow Q/c$, with 
a real constant $c$. This merely corresponds to describing CTM with the exchange of the variables from the beginning, 
and does not correspond to an essential change of the framework of CTM. 
Thus, we can define the Hamiltonian constraint of mCTM to be
\[
{\cal H}^{1}_a=\alpha_1 Q_{abc} P_{bde}P_{cde} +\alpha_2 Q_{abc}Q_{bde}Q_{cde}+\alpha_3 P_{abb}+\alpha_4 Q_{abb},
\label{eq:ham1}
\]
where the parameters $\alpha_i$ are real, and satisfy $\alpha_1=-\alpha_2 c^4, \alpha_3=-\alpha_4 c^2$. 
The Hamiltonian constraint \eq{eq:ham1}
is invariant under the time-reversal transformation, $Q\rightarrow Q,\ P\rightarrow -P$,
only when $\alpha_3=0$, which corresponds to $\lambda=0$ in CTM.
This restriction might be acceptable, because, as discussed in \cite{Sasakura:2015pxa},
the $\lambda$ term causes a difficulty of violating the locality of the formal continuum limit. 
However, for the generality of the following discussions, we consider the general values of $\alpha_i$, not necessarily 
restricted by the correspondence to CTM.

The quantization of ${\cal H}^1$ can be performed by promoting $Q,P$ to the quantized variables, $\hQ,\hP$,
as in Section~\ref{sec:review}. 
To obtain a quantized hamiltonian, the normal ordering in the first term of \eq{eq:ham1} must be fixed.
Normally, there are ambiguities, but, in the present case, the quantized hamiltonian is uniquely determined
from the classical one by imposing the $O(N)$ covariance and the hermiticity of the Hamiltonian.
Firstly, one can always reorder the first term into the symmetric form $\hP_{bde} \hQ_{abc} \hP_{cde}$,
which is hermite.
Then, in this process of reordering, a number of terms of $i\hat P_{\ldots}$ will be generated, but
they must be collected into the covariant form $i P_{abb}$, because of the requirement of the $O(N)$ covariance. 
Then, the hermiticity of the Hamiltonian requires this term to vanish, and hence the quantized Hamiltonian is uniquely 
determined to be
\[
\hat{\cal H}^1_a= \alpha_1  \hP_{bde} \hQ_{abc}\hP_{cde} +\alpha_2 \hQ_{abc}\hQ_{bde}\hQ_{cde}+\alpha_3 \hP_{abb}
+\alpha_4 \hQ_{abb},
\label{eq:qham1}
\]
where all the $\alpha_i$ are real as in the classical case.

The presence of the cubic term of $Q$ in \eq{eq:ham1} and \eq{eq:qham1} makes it difficult to 
obtain the explicit solutions to the constraints, which will be studied in later sections. In fact, one 
can transform them so that the maximum orders of each $Q$ and $P$ are quadratic, as follows.
When the relative sign between $\alpha_1$ and $\alpha_2$ is minus, namely in the case which corresponds to 
the original CTM, ${\cal H}^1$ can be transformed to the following form,
\[
{\cal H}^2_a=\beta_1 P_{abc} P_{bde}Q_{cde}
+\beta_2 Q_{abc} Q_{bde}P_{cde}+\beta_3 P_{abb} +\beta_4Q_{abb}
\label{eq:ham2}
\]
with real $\beta_i$, by an $SL(2,R)$ transformation,
\[
\left(
\begin{array}{c}
Q'_{abc} \\
P'_{abc}
\end{array}
\right)
=
\left(
\begin{array}{cc}
z_1 & z_2 \\
z_3 & z_4
\end{array}
\right)
\left(
\begin{array}{c}
Q_{abc} \\
P_{abc}
\end{array}
\right),
\ \ \ z_1 z_4-z_2 z_3=1,\ z_i\in \hbox{R},
\label{eq:sl2r}
\]
which keeps the fundamental Poisson bracket \eq{eq:poisson}.
The details are given in Appendix~\ref{app:trans1to2}.

The quantization of ${\cal H}^2$ can be done in a similar manner as $\hat {\cal H}^1$. 
The Hamiltonian takes the form similar to \eq{eq:ham2} as 
\[
\hat {\cal H}^2_a=\beta_1 \hP_{abc} \hP_{bde}\hQ_{cde}+\beta_2\hQ_{abc} \hQ_{bde}\hP_{cde}
+\beta_3 \hP_{abb} +\beta_4 \hQ_{abb},
\label{eq:qham2}
\]
where $\beta_1,\beta_2$ are real. 
The difference is the existence of the normal ordering terms as in \eq{eq:qhj} of CTM,
which appear as the imaginary parts of $\beta_3,\beta_4$.
The hermiticity condition determines
\[
\frac{1}{\beta_1}\hbox{Im}[\beta_3]=-\frac{1}{\beta_2}\hbox{Im}[\beta_4]=\frac{(N+2)(N+3)}{2}.
\label{eq:imbeta}
\]

The expressions of ${\cal H}^2$ and $\hat {\cal H}^2$ are very similar to those of CTM in \eq{eq:ctmconstraints}
and \eq{eq:qhj}. 
Actually, the only difference is the coexistence of the two cubic terms, and CTM can be 
obtained by taking one of $\beta_1$ or $\beta_2$ to zero\footnote{Actually, the constraint algebra of CTM is
invariant under the exchange of $Q,P$, meaning the equivalence of the two options.}.
Therefore, an advantage of the expressions of ${\cal H}^2,\hat{\cal H}^2$ 
over ${\cal H}^1,\hat {\cal H}^1$ is that one would be able to
import some of the known results from the original CTM 
with appropriate modifications related to the additional term.  
In fact, in the following sections, we will discuss the solutions to the constraints
and the relation with the network dynamics by using 
${\cal H}^2,\hat{\cal H}^2$ rather than ${\cal H}^1,\hat {\cal H}^1$.

When the relative sign between $\alpha_1$ and $\alpha_2$ is positive, 
it is not possible to transform ${\cal H}^1$ 
into the form of \eq{eq:ham2} by an $SL(2,R)$ transformation (See Appendix~\ref{app:trans1to2}).
However, in the quantized case, one can transform it to another form resembling $\hat {\cal H}^2$: 
\[
\hat {\cal H}_{a}^3=\gamma_1 A^\dagger_{abc} A^\dagger_{bde}A_{cde}+\gamma_2 A^\dagger_{cde} A_{bde} A_{abc}
+\gamma_3 A^\dagger_{abb} + \gamma_4 A_{abb}
\label{eq:qham3}
\] 
Here, $A^\dagger,A$ are creation-annihilation operators satisfying
$[A_{abc},A^\dagger_{def}]=\sum_{\sigma} \delta_{a\sigma_d}
\delta_{b\sigma_e}\delta_{c\sigma_f}$.
This rewriting is obvious, if one notices that the cubic part of ${\cal H}^1$ has a 
form like $\sim Q_{abc} H_{bc}^{harm}$, where
$H^{harm}$ has the form of the hamiltonian of coupled 
harmonic oscillators. The hermiticity of $\hat {\cal H}^3$ 
requires $\gamma_1^*=\gamma_2,\ \gamma_3^*=\gamma_4$. 

We may regard all the above three Hamiltonians as the definitions of mCTM in a loose sense. 
In a strict sense, CTM can only be derived from those through the Hamilton-Jacobi procedure 
with some restrictions on the parameters.

\section{Non-trivial solutions in the quantum case}
\label{sec:quantum}

In the quantum case, an important problem is to obtain a physical state (or a physical wave function),
which must satisfy
\[
\begin{split}
\hat {\cal H}_a \Psi&=0,\\
\hat {\cal J}_{ab} \Psi&=0.
\end{split}
\label{eq:HJeq}
\]
What is simple in the quantum case is that all the secondary constraints are automatically satisfied by $\Psi$,
if \eq{eq:HJeq} is satisfied:
\[
[\hat {\cal H}_a,\hat {\cal H}_b]\Psi=0, \hbox{ and so on.}
\]
Therefore, it is enough to solve \eq{eq:HJeq}. 

An important question is whether such physical states exist with 
non-trivial properties, or not. If the solutions did not exist or were not 
interesting, the theory would be vacuous or uninteresting. 
In the following subsections, we will answer this question by 
explicitly giving some exact solutions first for $N=1$ and also
for general $N$ with restricted parameter values. 
These solutions seem non-trivial and interesting.
Throughout this section, we assume $\beta_1,\beta_2,\gamma_1,\gamma_2 \neq 0$ to exclusively consider mCTM. 

\subsection{N=1}
\label{sec:n=1}
In the case $N=1$, \eq{eq:HJeq} gives just one equation,
\[
\hat{\cal H}_1 \Psi=0.
\]
This can be solved directly by solving the ordinary second-order 
differential equation for a wave function. From $\hat {\cal H}^2$ 
in \eq{eq:qham2}, the differential equation is given by
\[
\left[-\beta_1 \frac{d^2}{dq^2}q -i \beta_2 q^2 \frac{d}{dq} 
-i \beta_3 \frac{d}{dq} +\beta_4 q \right] \Psi(q)=0,
\]
where we have taken a representation, $Q=q,P=-i\frac{d}{dq}$ (Here, we have ignored a factor of 6, which appears if we 
take \eq{eq:poisson} literally, for the simplicity of the expression.). 
After a change of variable, $x=-i \beta_2 q^2/(2\beta_1)$, 
one can obtain the (Kummer's) confluent hypergeometric differential equation,
\[
x \Psi''+(A_2-x)\Psi'-A_1 \Psi=0,
\label{eq:conful}
\]
where $'$ denotes the derivative with respect to $x$, and 
\[
\begin{split}
A_1&=\frac{i \beta_4}{2 \beta_2}, \\
A_2&=\frac{3}{2}+\frac{i \beta_3}{2\beta_1}.
\end{split}
\label{eq:valuesA}
\]
When $A_2$ is not an integer, the general solution to \eq{eq:conful} is given by a linear combination,
\[
\Psi(x)=B_1 F(A_1,A_2;x)+B_2 x^{1-A_2} F(A_1-A_2+1,2-A_2;x),
\label{eq:gensol}
\]
where $B_i$ are arbitrary numerical coefficients and $F(A_1,A_2;x)$ 
denotes the (Kummer's) confluent hypergeometric function. 
See Appendix~\ref{app:n=1wavefn} for some details.
When $A_2$ is an integer, the solutions can be obrtained by taking some appropriate limits 
of \eq{eq:gensol}.

\begin{figure}
\center{
\includegraphics[scale=.4]{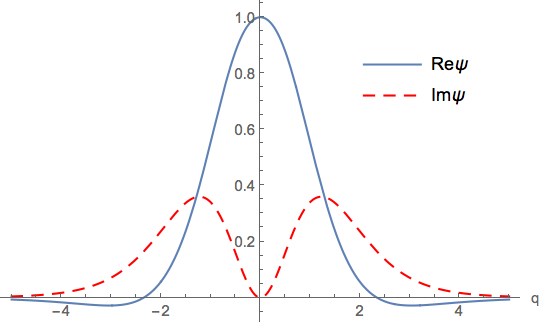}
\hspace{1cm}
\includegraphics[scale=.4]{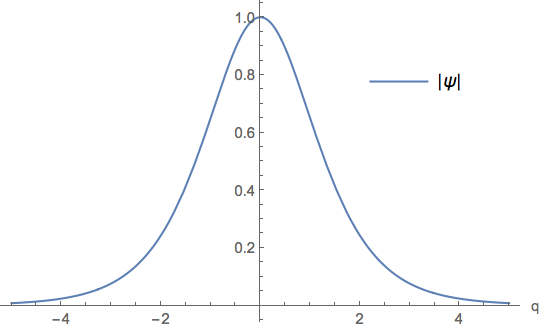}
}
\caption{The $N=1$ physical wave function for $\beta_1=\beta_2=\hbox{Re}[\beta_3]=\hbox{Re}[\beta_4]=1$
with $B_1=1$ and $B_2$ satisfying the condition \eq{eq:cancelsecond}. $\hbox{Im}[\beta_3]$ and $\hbox{Im}[\beta_4]$ 
are taken to be \eq{eq:imbeta} with $N=1$.
The left figure plots the real and imaginary parts of the wave function, 
and the right one plots the absolute value. }
\label{fig:waven=1}
\end{figure}

The physical meaning of a wave function of a spacetime is not clear, in the sense
that we cannot yet give a probabilistic interpretation to it, as is possible in the 
context of usual quantum mechanics. Currently we are lacking 
a clear intuitive understanding of it as we cannot yet give a 
meaning to the norm and phase of the wave-function. 
We also don't know the right set of boundary conditions 
that need to be imposed. This is partially because we lack 
a physical understanding of the wave-function. However, 
following the knowledge of usual quantum mechanics, it is 
interesting to investigate how the wave-function behaves 
when decaying boundary conditions at $|q|\rightarrow \infty$ 
are imposed. This is a reasonable assumption which we 
adopt in this paper. Then from the formula of the asymptotic 
behavior of the confluent geometric function (See Appendix~\ref{app:n=1wavefn}), 
one can find that, if 
\[
B_1 \frac{\Gamma(A_2)}{\Gamma(A_1)}+B_2 \frac{\Gamma(2-A_2)}{\Gamma(A_1-A_2+1)}=0,
\label{eq:cancelsecond}
\]
is satisfied, the asymptotically divergent part of the wave function \eq{eq:gensol} vanishes, 
and the asymptotic behavior reduces to $\sim q^{-6}$. 
To derive this, the explicit values of the imaginary parts of $\beta_3,\beta_4$ 
in \eq{eq:imbeta} with $N=1$ have been used.  See Appendix~\ref{app:n=1wavefn} for more details.
As an example, the wave function for $\beta_1=\beta_2=\hbox{Re}[\beta_3]=\hbox{Re}[\beta_4]=1$
is plotted in Figure~\ref{fig:waven=1}.

Let us next consider solving the physical state condition,
\[
\hat {\cal H}^3 |\Psi \rangle=0,
\label{eq:h3psi}
\]
where $\hat {\cal H}^3$ is given in \eq{eq:qham3}. 
Let us assume that the state can be expressed in terms of an expansion in a Fock space:
\[
 |\Psi \rangle=\sum_{n=0}^\infty c_n (A^\dagger)^n |0 \rangle,
 \label{eq:n=1physstate}
\]
where $A^\dagger,A$ represent the only oscillator in $N=1$, $|0\rangle$ denotes the Fock vacuum,
and $c_n$ are the coefficients to be determined.
It is obvious that, by considering a corresponding ``wave function", 
\[
\Psi(q)=\sum_{n=0}^\infty c_n q^n,
\label{eq:psiq}
\]
with a variable $q$, the condition \eq{eq:h3psi} is the same as
\[
\left[ \gamma_1 q^2 \frac{d}{dq} +\gamma_2 q \frac{d^2}{dq^2}
+\gamma_3 q+\gamma_4 \frac{d}{dq} \right]\Psi(q)=0.
\]
The solution is again given by the confluent hypergeometric function as \eq{eq:gensol} with 
\[
\begin{split}
x&= -\frac{\gamma_1}{2 \gamma_2}q^2,\\
A_1&=\frac{\gamma_3}{2 \gamma_1},\\
A_2&=\frac{1}{2} + \frac{\gamma_4}{2 \gamma_2}.
\end{split}
\]

When $A_2\neq 0,-1,-2,\ldots$, the first term of \eq{eq:gensol} (by setting $B_2$=0)
gives a solution consistent with the perturbative expression \eq{eq:psiq}.  
From \eq{eq:expF} in Appendix~\ref{app:n=1wavefn}, the explicit expression is given by
\[
\begin{split}
&c_{2n}=\frac{(A_1)_n}{(A_2)_n n!}\left(-\frac{\gamma_1}{2 \gamma_2} \right)^n c_0, \\
&c_{2n+1}=0,
\end{split}
\]
with integer $n \geq 0$.
Since $c_n\sim n^{const.} const.^n/n!$ for large $n$, the norm of the physical state \eq{eq:n=1physstate}, 
\[
\langle \Psi |\Psi \rangle=\sum_{n=0}^\infty |c_n|^2 n! , 
\]
is finite. 

When $A_2=\frac{1}{2},0,-\frac{1}{2},-1,\ldots$, the second term 
of \eq{eq:gensol} (by setting $B_1=0$) gives a 
solution\footnote{One can consider $A_2=1$, but this is degenerate with the first case.}. 
The perturbation starts from $c_{2-2A_2}$, and 
\[
\begin{split}
&c_{2n+2-2A_2}=\frac{(A_1-A_2+1)_n}{(2-A_2)_n n!} 
\left(-\frac{\gamma_1}{2 \gamma_2} \right)^n c_{2-2A_2}, \\
&c_{2n+3-2A_2}=0, 
\end{split}
\]
with integer $n\geq 0$. Similarly, the state can be shown to be normalizable.

\subsection{General $N$}
In this subsection, we will give an exact solution to \eq{eq:HJeq} valid for general $N$ 
by using an insight from the randomly connected tensor network (RCTN). The insight was also greatly useful
in our previous study of obtaining the exact physical wave functions of CTM \cite{Narain:2014cya}.
The generality of $N$ is remarkable, but, as we will see, there are 
two disadvantages compared to the $N=1$ solution 
obtained in Section~\ref{sec:n=1}.
One is that one of the parameters, $\beta_3$ or $\beta_4$, must be 
taken to be a specific value. The other is that the expression of the solution 
still contains complex-valued integrations over $N$ variables.
This makes it difficult to study even the qualitative properties of the solution.
Therefore, we will only check the non-triviality of the solution by taking special values 
of the parameters to make the expression simple enough.

Let us first recall some basics of RCTN. 
The partition function of RCTN with $n$ vertices (tensors) can be defined 
by \cite{Sasakura:2014zwa,Sasakura:2014yoa}
\[
Z_n(Q)=\int d^N\phi \ (Q\phi^3)^n e^{-\phi^2},
\label{eq:partRCTN}
\]
where we have used short-hand notations,
\[
\begin{split}
\int d^N\phi&= \prod_{a=1}^N \int_{-\infty}^\infty d\phi_a, \\
\phi^2&=\phi_a\phi_a,\\
Q\phi^3&=Q_{abc}\phi_a\phi_b\phi_c.
\end{split}
\]
By applying the Wick theorem for the Gaussian integration,
the partition function can be expressed as a summation over all the possible contractions of $n$ $Q$'s.
This defines the randomly connected tensor network of a rank-three tensor $Q$.
It is obvious that the partition function is invariant under the $O(N)$ transformation \eq{eq:ontransformation}.

We can also consider a ``grand" partition function of the system \cite{Sasakura:2014zwa} by
\[
Z_{\cal C}(Q)=\sum_{n=0}^\infty \frac{1}{n!} Z_n(Q)
=\int_{\cal C} d^N\phi \ e^{-\phi^2+Q\phi^3}.
\label{eq:grandz}
\]
Actually, the summation is divergent and ill-defined. 
This is reflected in the existence of the $\phi^3$ coupling term in the exponent of the integrand
in the rightmost expression.
The integration can be made well-defined by deforming the integration 
region of $\phi$ to an appropriate complex integration contour 
${\cal C}$ such that  ${\rm Re}[-\phi^2+Q\phi^3]\rightarrow -\infty$ for $|\phi|\rightarrow\infty$. 
The easiest way is to make the rotation $\phi_a=e^{i \pi/6} r_a$ 
with real $r_a$ \cite{Sasakura:2014zwa}. Then, the divergent summation 
in \eq{eq:grandz} can be understood as an asymptotic expansion of 
$Z_{\cal C}(Q)$ around $Q=0$.
In fact, in $N=1$, $Z_{\cal C}(Q)$ is essentially the Airy function. In this sense, \eq{eq:grandz}
defines a multi-integral generalization of the Airy function for general $N$.

Let us first consider the first equation of the physical state condition \eq{eq:HJeq}.
By employing the Hamiltonian $\hat{\cal H}^2$ in \eq{eq:qham2}, it is given by
\[
\left[-\beta_1 D_{abc} D_{bde}Q_{cde}-i \beta_2 Q_{abc} Q_{bde}D_{cde}
-i\beta_3 D_{abb} +\beta_4Q_{abb}
\right] \Psi(Q)=0,
\label{eq:state2bare}
\] 
where $\Psi(Q)$ is the physical wave function represented in $Q$, and 
$D$ denotes the derivation with respect to $Q$. It is defined by 
\[
D_{abc}Q_{def}=\sum_{\sigma} \delta_{a\sigma_d}\delta_{b\sigma_e}\delta_{c\sigma_f}.
\]
It is convenient to change the order of $Q$ and $D$ in the first term of \eq{eq:state2bare} as  
\[
\left[-\beta_1Q_{cde} D_{abc} D_{bde}-i \beta_2 Q_{abc} Q_{bde}D_{cde}-
\beta'_3 D_{abb} +\beta_4Q_{abb}
\right] \Psi(Q)=0,
\label{eq:state2}
\]
where $\beta'_3=(N+2)(N+3)\beta_1+i\beta_3$.

To obtain a solution to \eq{eq:state2},
let us consider the following assumption for the wave function, which is obtained by 
a slight generalization of \eq{eq:grandz},
\[
\Psi_{f,{\cal C}}(Q)=\int_{\cal C} d^N\phi\ f(\phi^2)\ e^{Q\phi^3},
\label{eq:psicfq}
\]
where $f(\phi^2)$ is a holomorphic function of $\phi^2$.
Then, the first term of \eq{eq:state2} can be computed as 
\[
\begin{split}
Q_{cde}D_{abc}D_{bde} \Psi_{f,{\cal C}}(Q)&=36 \int_{\cal C} d^N\phi\ Q_{cde} 
\phi_a \phi_b \phi_c \phi_b \phi_d \phi_e f(\phi^2) e^{Q\phi^3} \\
&=12  \int_{\cal C} d^N\phi\ \phi_a \phi_c \phi^2 f(\phi^2)\partial_c e^{Q\phi^3} \\
&=-12  \int_{\cal C} d^N\phi\ \phi_a\left((N+3)\phi^2 f(\phi^2)+2 (\phi^2)^2f'(\phi^2)\right) e^{Q\phi^3},
\end{split}
\label{eq:h2first}
\]
where $\partial_a=\frac{\partial}{\partial \phi_a}$, and we have 
performed a partial integration for $\phi$ on the assumption that the 
boundary terms are ignorable. This assumption can be justified, if the integrand damps 
fast enough at infinity of ${\cal C}$, ${\cal C}$ is a cycle with no 
boundaries\footnote{An explicit example with a cycle 
was previously considered in \cite{Narain:2014cya}.},
or their mixtures.

Similarly, the second term in \eq{eq:state2} reduces to 
\[
\begin{split}
Q_{abc}Q_{bde}D_{cde}  \Psi_{f,{\cal C}}(Q)
&=6 \int_{\cal C} d^N\phi\ Q_{abc}Q_{bde}\phi_c \phi_d \phi_e f(\phi^2) e^{Q\phi^3} \\
&=2 \int_{\cal C} d^N\phi\ Q_{abc}\phi_c f(\phi^2) \partial_b e^{Q\phi^3}\\
&=-2 \int_{\cal C} d^N\phi\ \left( Q_{abb} f(\phi^2) +2 Q_{abc} \phi_b \phi_c f'(\phi^2) \right)e^{Q\phi^3} \\
&=-2 Q_{abb} \Psi_{f,{\cal C}}(Q)-\frac{4}{3} \int_{\cal C} d^N\phi\ f'(\phi^2) \partial_a e^{Q\phi^3} \\
&= -2 Q_{abb} \Psi_{f,{\cal C}}(Q)+\frac{8}{3} \int_{\cal C} d^N\phi \ \phi_a f''(\phi^2) e^{Q\phi^3}.
\end{split}
\label{eq:h2second}
\]
And, the third is obtained as
\[
D_{abb}  \Psi_{f,{\cal C}}(Q)&=6 \int_{\cal C} d^N\phi\ \phi_a \phi^2 f(\phi^2) e^{Q\phi^3}.
\label{eq:h2third}
\]

By putting \eq{eq:h2first}, \eq{eq:h2second} and \eq{eq:h2third} into \eq{eq:state2}, 
we obtain the following satisfactory conditions for solving \eq{eq:state2}:
\[
&2i\beta_2+\beta_4=0,
\label{eq:be2be4}\\
&12 \beta_1 \left((N+3)z f(z)+2 z^2f'(z)\right)-\frac{8 i \beta_2}{3} f''(z)-6 \beta'_3 z f(z)=0.
\label{eq:diffgeneralN}
\]
The first equation comes from the vanishing of $Q_{abb}  
\Psi_{f,{\cal C}}$, and the second from the vanishing of 
$\int d^N\phi\ \phi_a e^{Q\phi^3}(\cdots)$.The solution to 
\eq{eq:diffgeneralN} is again given by \eq{eq:gensol} with
\[
\begin{split}
x&=-\frac{3  i \beta_1 z^3}{\beta_2}, \\
A_1&=-\frac{\beta'_3}{12 \beta_1}+\frac{N+3}{6}
=-\frac{i \beta_3}{12 \beta_1}-\frac{N(N+3)}{12
}, \\
A_2&=\frac{2}{3}.
\end{split}
\label{eq:solcond}
 \]
It is obvious that, when we use $P$-representation instead of $Q$, 
the roles of $\beta_i$ are interchanged in \eq{eq:be2be4} and \eq{eq:solcond}.

It would be interesting to study the properties of the wave function we have obtained. However,
because of the integral form \eq{eq:psicfq} with a confluent hypergeometric function $f$, 
it does not seem easy even to check whether \eq{eq:psicfq} gives a physically interesting wave function or not.
Because of that, we will only consider a simple case with a specific choice of the values of $\beta_i$
to show that it is indeed non-trivial.
The values of $\beta_i$ we will take are not included in the range of physical values, 
namely, $\beta_1,\beta_2$ being real and $\beta_3,\beta_4$ taking the imaginary parts given in \eq{eq:imbeta}. 
However, we would be able to expect that the physical wave function would remain non-trivial by
the analytic continuation from the unphysical to the physical values.
It is obvious that more detailed study in future is necessary.

As a simple case, let us take the values of $\beta_i$ in \eq{eq:solcond} so that
\[
\begin{split}
x&=z^3, \\
A_1&=A_2=\frac{2}{3}.
\end{split}
\label{eq:simplevalues}
\]
One can easily check that the corresponding values of $\beta_i$ are not physical.
In this case, a solution to the confluent hypergeometric differential 
equation is given by $f(z)=e^{z^3}$. Then, the physical wave function is given by
\[
\begin{split}
\Psi_{f,\cal C}(Q)&=\int_{\cal C} d^N\phi\ e^{(\phi^2)^3+Q\phi^3}.
\end{split}
\label{eq:simpleint}
\]
The simplest choice of a contour ${\cal C}$ for the convergence is to 
take it along the imaginary axes. Then, with
the replacement $\phi\rightarrow i \phi$, the wave function can be expressed as 
 \[
\begin{split}
\Psi_{f,{\rm Im}}(Q)&=\int_{R^N} d^N\phi\ e^{-(\phi^2)^3-i Q\phi^3}
\end{split}
\label{eq:simpleintR}
\]
with the integration over $R^N$. 
It is obvious that the integral \eq{eq:simpleintR} is convergent for any real values of $Q$, and,
the damping of the integrand for large $|\phi|$ is fast enough to justify the partial integrations 
assumed for the derivation of the solution. 
One can easily show that, due to the oscillation,
the wave function damps at large $|Q|$ as $\Psi_{f,{\rm Im}}(Q)\sim Q^{-N/3}$.
However, since the dimension of the configuration space of $Q$ is $N(N+1)(N+2)/6$,
the norm of the wave function is not finite. 

In principle, one can improve the situation by taking a contour ${\cal C}$ 
which passes through a non-zero saddle point at $\phi\neq 0$.
Here, ${\cal C}$ must be assumed to be taken so that the real part of the exponent of the integrand in 
\eq{eq:simpleint} decreases as leaving away from the saddle point \cite{singularity}.
Let us assume that this can be done to infinity to get a convergent integration. Then,
by performing a rescaling $\phi\rightarrow |Q|^{1/3} \phi$, 
the argument in \eq{eq:simpleint} becomes $|Q|^2( -(\phi^2)^3+i Q\phi^3/|Q| )$.
With this expression, a non-trivial saddle point at  $\phi\neq 0$ does not depend on the overall scaling of $Q$, 
and therefore the wave function will damp exponentially in $|Q|^2$. This will give a normalizable wave function.

In the next, we want to solve for the physical wave function for $\hat {\cal H}^3$ in \eq{eq:qham3}:
\[
\hat{\cal H}_a^3 |\Psi\rangle=0.
\]
As an assumption, we similarly consider 
\[
|\Psi\rangle =\int_{\cal C} d^N\phi\ f(\phi^2)\, e^{A^\dagger \phi^3}|0\rangle,
\]
where $|0\rangle$ is the Fock vacuum, 
and $A^\dagger \phi^3=A_{abc}^\dagger \phi_a \phi_b \phi_c$.
This expression should be interpreted in the perturbative expansion of the exponential function. For this to be 
well-defined, all the integrations of $f(\phi^2)$ multiplied by polynomials of $\phi$ must converge: 
\[
\int_{\cal C} d^N\phi\ f(\phi^2) \, \phi_{a_1} \phi_{a_2}\cdots  \phi_{a_n} =\hbox{finite}.
\label{eq:intfppp}
\]
In addition, we assume that the partial integration over $\phi$ can be justified with no boundary contributions
to use the same trick in the case of $\hat{\cal H}^2$.
Then, we obtain
\[
&-2 \gamma_1+\gamma_3=0,\\
&\frac{8}{3} \gamma_1 f''(z)-12 \gamma_2 \left( (N+3)z f(z)+2 z^2 f'(z)\right)+6 \gamma_4 z f(z)=0.
\]
The solution is again obtained by \eq{eq:gensol} with 
\[
\begin{split}
x&=\frac{3 \gamma_2 z^3}{\gamma_1}, \\
A_1&=-\frac{\gamma_4}{12 \gamma_2}+\frac{N+3}{6}, \\
A_2&=\frac{2}{3}.
\end{split}
\] 

The finiteness condition \eq{eq:intfppp} can be satisfied, if we consider an $f(z)$ which 
damps faster than any polynomials. From the asymptotic expansion \eq{eq:asymF},
one can see that this can be achieved by cancelling the polynomial 
part of the asymptotic expansion. This can be realized by tuning $B_1,B_2$ as
\[
B_1 \frac{\Gamma(A_2)}{\Gamma(A_2-A_1)}
+(-1)^{1-A_2}B_2 \frac{\Gamma(2-A_2)}{\Gamma(1-A_1)}=0.
\]  
Then, the asymptotic behavior of $f(z)$ at infinity is 
$\sim \exp (3 \gamma_2 z^3/\gamma_1)$, and \eq{eq:intfppp} can 
be satisfied by taking ${\cal C}$ so that $\exp (3 \gamma_2 (\phi^2)^3/\gamma_1)$ damps
exponentially in the infinity.

We finally want to discuss the constraint 
\[
\hat {\cal J}_{ab} \Psi=0.
\]
For the wave function \eq{eq:psicfq}, we obtain
\[
\begin{split}
\hat {\cal J}_{ab} \Psi&=\left(Q_{acd}D_{bcd}-Q_{bcd}D_{acd}\right) \int_{\cal C} d^N\phi\ f(\phi^2)e^{Q\phi^3} \\
&=6 \int_{\cal C} d^N\phi\ \left(Q_{acd}\phi_b \phi_c \phi_d-Q_{bcd}\phi_a \phi_c \phi_d\right) f(\phi^2) e^{Q\phi^3}\\
&=2 \int_{\cal C} d^N\phi\  f(\phi^2)\left(\phi_b \partial_a -\phi_a \partial_b \right) e^{Q\phi^3}\\
&=-2 \int_{\cal C} d^N\phi\  \left( \partial_a (f(\phi^2)\phi_b) - \partial_b (f(\phi^2)\phi_a) \right) e^{Q\phi^3}\\
&=0,
\end{split}
\]
where we have assumed the validity of the partial integration 
with no boundary contributions. Therefore, under the assumption of the validity of partial integrations, 
the ${\cal J}$-constraint is automatically satisfied by \eq{eq:psicfq}.

\section{Non-trivial solutions in the classical case}
\label{sec:classical}
Comparing to the quantum case, the problem of finding the classical 
solutions to the constraints of mCTM is much more complicated. 
The reason is that, unlike CTM which possesses a closed algebra \eq{eq:ctmconstraintalgebra}
of the ``Hamiltonian" and ``momentum" constraints, 
we do not know a full set of the classical constraints of mCTM.
We tried to find a full set by explicitly computing a few of the secondary constraints 
starting from the Hamiltonian ${\cal H}^2$, but the expressions 
we obtained were too complicated to find such a set. 
Therefore, in this section, rather than finding such a full set of constraints, we will present some classical solutions 
which can be proven to satisfy all the secondary constraints (and the primary ones of course). 
Starting with the $N=1$ case which can be solved explicitly,
we will give two other examples valid for general $N$, 
and another numerical example valid exclusively for $N=2$.

\subsection{Classical solutions for $N=1$}
\label{sec:classsolN=1}
In the $N=1$ case, mCTM has only one ``Hamiltonian" constraint \eq{eq:ham2},
with no ``momentum" constraints and no secondary constraints. 
We assume $\beta_1,\beta_2\neq0$ to exclusively consider mCTM.
Solving the Hamiltonian constraint for $P$, we obtain
\[
P=\frac{-(\beta_2 Q^2+\beta_3)\pm\sqrt{ (\beta_2 Q^2+\beta_3)^2-4 \beta_1 \beta_4 Q^2}}{2 \beta_1 Q}.
\label{eq:solP}
\]
Then, we find
\[
\begin{split}
\frac{d}{dt}Q&=\{Q,H\}\\
&=\pm \sqrt{(\beta_2 Q^2+\beta_3)^2-4 \beta_1 \beta_4 Q^2},
\end{split}
\label{eq:eqQ}
\]
where the lapse has been taken to be $n=1$ (We have taken $\{Q,P\}=1$, ignoaring a factor of 6 in \eq{eq:poisson}
for the simplicity of the expression.).

\begin{figure}
\begin{center}
\includegraphics[scale=.25]{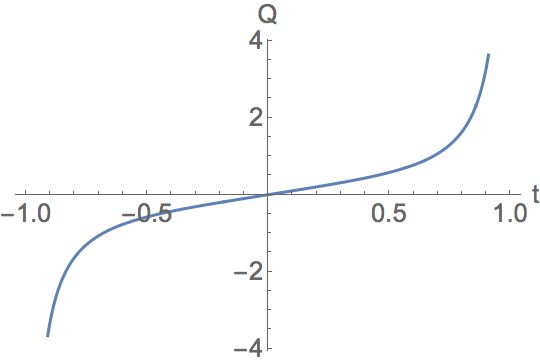}
\includegraphics[scale=.25]{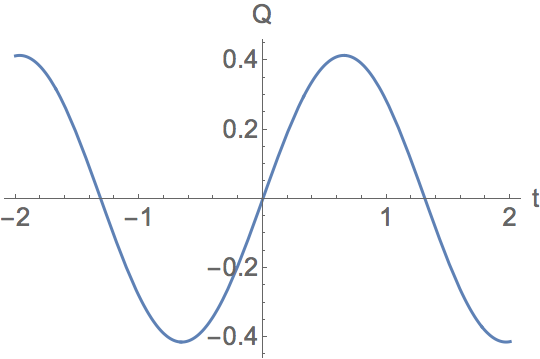}
\includegraphics[scale=.25]{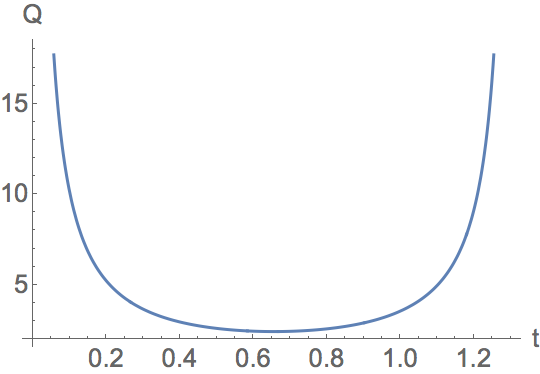}
\caption{
The classical solution is drawn with \eq{eq:solQ}.
In the leftmost figure, a blow-up solution is shown for the case (i) with $(\beta_1,\beta_2,\beta_3,\beta_4)=(1,3,1,1)$.
The blow-up occurs at $t=i K(\sqrt{1-k^2})/Q_2\beta_2$, where $K(k)$ is the elliptic integral of the first kind. 
In the middle figure, an oscillatory solution is shown for the case 
(ii) with  $(\beta_1,\beta_2,\beta_3,\beta_4)=(1,1,1,2)$. The period is $4 K(k)/Q_2 \beta_2$.
In the rightmost figure, a bounce solution is shown for the same case. The integration constant is taken as
$c_0=i K (\sqrt{1-k^2})$ in \eq{eq:solQ}.
The blow up occurs at $t=0,\ 2 K(k)/Q_2 \beta_2$.
In the case (iii), the turning points in the middle and the rightmost figures are elongated to infinite time intervals, 
since $K(k)\rightarrow \infty$ for $k=\frac{Q_1}{Q_2}\rightarrow 1$. 
}
\label{fig:drawQ}
\end{center}
\end{figure}

The differential equation \eq{eq:eqQ} can be solved by the Jacobi elliptic functions. The solution is given by
\[
Q(t)&=Q_1 \hbox{sn} (Q_2  \beta_2\,  t+c_0,k),\ \ \ k=\frac{Q_1}{Q_2},
\label{eq:solQ}
\]
where $c_0$ is an integration constant, and $Q_1^2, Q_2^2$ are the two solutions to the quadratic equation 
$(\beta_2 x+\beta_3)^2-4 \beta_1 \beta_4 x=0$.
However, 
it is easier to obtain the qualitative behaviors of the solutions by studying the positive regions of the argument 
of the square root in \eq{eq:eqQ}, rather than using the exact expression \eq{eq:solQ}.
Since our purpose is not a full classification, let us assume $\beta_3\neq 0$ for simplification.
Then, the behaviors of the classical solutions can be classified as follows:
\begin{itemize}
\item[(i)] $\beta_1\beta_4(\beta_2 \beta_3-\beta_1\beta_4)>0$ or $\beta_2\beta_3-2 \beta_1 \beta_4 \geq 0$ \\
For any value of $Q$, the argument of the square root in \eq{eq:eqQ} is positive. 
At large $|Q|$, the righthand side of \eq{eq:eqQ} is $\sim Q^2$. Therefore, 
the solution blows up in a finite time. 
\item[(ii)] $\beta_1\beta_4(\beta_2 \beta_3-\beta_1\beta_4)<0$ and $\beta_2\beta_3-2 \beta_1 \beta_4 < 0$\\
The argument of the square root becomes non-negative in the three regions, 
$Q\leq -Q_2,\ -Q_1 \leq Q \leq Q_1,\ Q_2\leq Q$,  where we take $0<Q_1<Q_2$ without loss of generality.
In the two outer regions, the solution blows up in a finite time (possibly after a bounce at $\pm Q_2$),
as in the case (i). In the middle region, the solution oscillates forever between $\pm Q_1$.
\item[(iii)] $\beta_1\beta_4(\beta_2 \beta_3-\beta_1\beta_4)=0$ and $\beta_2\beta_3-2 \beta_1 \beta_4 < 0$\\
This is the degenerate case with $Q_1=Q_2$. Depending on the initial direction, the solution
asymptotically approaches $Q_1=Q_2$, or blows up in a finite time.
\end{itemize}
In Figure~\ref{fig:drawQ}, the solutions are explicitly drawn for some example cases by using the exact expression \eq{eq:solQ}. 

\begin{figure}
\center{
\includegraphics[scale=.4]{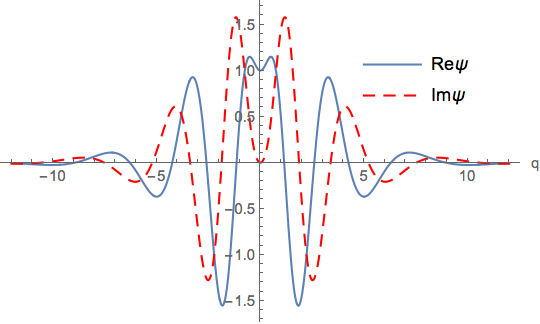}
\hspace{1cm}
\includegraphics[scale=.4]{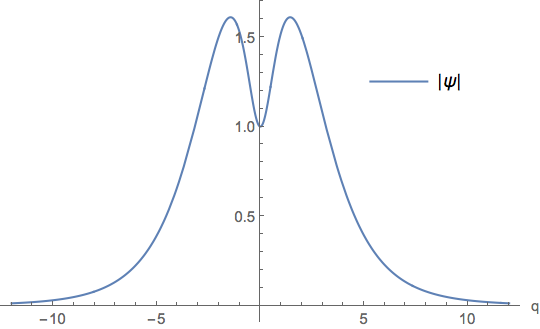}
}
\caption{The $N=1$ physical wave function for $\beta_1=\beta_2=\hbox{Re}[\beta_3]=1$, $\hbox{Re}[\beta_4]=-10$
with $B_1=1$ and $B_2$ satisfying the condition \eq{eq:cancelsecond}. $\hbox{Im}[\beta_3]$ and $\hbox{Im}[\beta_4]$ 
are taken to be \eq{eq:imbeta} with $N=1$.
The left figure plots the real and imaginary parts of the wave function, 
and the right one plots the absolute value. }
\label{fig:caseii}
\end{figure}

In all the above three cases, we encounter the solutions which diverge in finite times. 
The system encounters singularities after finite times, which 
would question the validity of the classical treatment of mCTM 
and would rather imply the necessity of quantization.
In fact, in the quantum case, no serious troubles seem to exist: for all those three cases, 
we can obtain normalizable wave functions by the condition \eq{eq:cancelsecond}. 
Figure~\ref{fig:waven=1} corresponds to the case (iii), and the case (ii) has more or less similar profiles
of wave functions.
In case (i), as seen in Figure~\ref{fig:caseii},
interesting oscillatory patterns appear in a region around $Q=0$, where the momentum \eq{eq:solP}
becomes relatively large.
As a summary, the wave functions have moderate behaviors with
more or less similar profiles confined around $Q=0$ possibly with some oscillations, while 
the classical solutions have detailed structures depending on the parameters 
and suffer from divergences.
It seems that mCTM prefers the quantum mechanical treatment rather than the classical one.

\subsection{Solutions with general $N$ for $\beta_3/\beta_1=\beta_4/\beta_2$}
If $\beta_3/\beta_1=\beta_4/\beta_2$, 
there exist a finite set of constraints which assures that the solutions 
to them satisfy all the secondary constraints.
Let us introduce a parameter $\beta=\beta_3/\beta_1=\beta_4/\beta_2$. Then,  
the Hamiltonian constraint ${\cal H}^2$ can be rewritten in the form,
\[
\begin{split}
{\cal H}^2_a&=\beta_1 P_{abc} P_{bde} Q_{cde} 
+\beta_2 Q_{abc}Q_{bde}P_{cde} +\beta_3 P_{abb} +\beta_4 Q_{abb} \\
&=2 \left( \beta_1 P_{abc}+\beta_2 Q_{abc} \right) \ctJ_{bc},
\end{split}
\] 
where
\[
{\cal \tilde J}_{ab}= \frac{1}{4} \left( P_{acd}Q_{bcd}+P_{bcd}Q_{acd} + 2 \beta  \delta_{ab} \right).
\label{eq:deftildeJ}
\]
Here, $\ctJ$ is a symmetric partner of $\cal J$ in \eq{eq:ctmconstraints}, 
and forms the $gl(N)$ Lie algebra with $\cJ$:
\[
\begin{split}
\{ \ctJ (\tilde \eta^1), \ctJ(\tilde \eta^2) \}&=\cJ[\tilde \eta^1,\tilde \eta^2]),\\
\{\cJ(\eta),\ctJ (\tilde \eta) \} &= \ctJ([\eta,\tilde \eta]),\\
\{\cJ(\eta^1),\cJ (\eta^2) \} &= \cJ([\eta^1,\eta^2]),
\end{split}
\label{eq:glnalg}
\] 
where $\ctJ(\tilde \eta)=\tilde\eta_{ab} \ctJ_{ab} $ with $\tilde \eta_{ab}=\tilde \eta_{ba}$.
Because of this algebraic closure, all the secondary constraints generated by 
$\{\cH^2_{a_1},\{\cH^2_{a_2},\{\cdots,\cH^2_{a_m}\}\cdots\}$
can be expressed as linear combinations of $\ctJ$ and ${\cJ}$, 
with coefficients depending on $P$ and $Q$. 
Therefore, the solutions to $\ctJ_{ab}=\cJ_{ab}=0$ satisfy all the primary and secondary constraints. 

One can set an initial configuration satisfying $\ctJ_{ab}=\cJ_{ab}=0$, 
and solve the classical equation of motion for the Hamiltonian ${\cal H}^2$
to obtain a classical trajectory.
Then, the constraints are kept being satisfied over the classical trajectory, 
since all the secondary constraints are satisfied by the initial configuration.
In general, such trajectories have various behaviors, because the lapse $n_a$ can be taken arbitrary.
This makes it difficult to tell the general properties of the classical trajectories,
but an important common thing is that the constraints, $\ctJ,\cJ=0$, 
do not prohibit $Q,P$ from diverging. In fact, from short numerical study, 
one can find that classical trajectories easily diverge in finite times, similarly to the $N=1$ case.   
This would question the validity of the classical treatment of mCTM, as in the $N=1$ case.
      
\subsection{Another kind of solutions with general $N$ for $\beta_3/\beta_1=\beta_4/\beta_2$} 
There are another kind of classical solutions to all the constraints, if $\beta_3/\beta_1=\beta_4/\beta_2$. 
By performing a rescaling $Q\rightarrow r Q,\ P\rightarrow P/r$ with real $r$, which keeps the fundamental Poisson bracket 
\eq{eq:poisson}
unchanged, one can take $\beta_1=\beta_2,\ \beta_3=\beta_4$ without loss of generality.
In this case, the Hamiltonian is symmetric under the exchange of $Q,P$:
\[
\cH^2_{a}(Q,P)=\cH^2_{a}(P,Q).
\label{eq:symh2}
\]
Another obvious property is 
\[
\hbox{Order}(\cH^2)=\hbox{odd},
\label{eq:orderh2}
\]
which means that each term of $\cH^2$ contains an odd number 
of $Q$ or $P$ in total (one or three in the present case). Because of the two 
properties \eq{eq:symh2} and \eq{eq:orderh2}, $\cH^2$ can be schematically represented as 
\[
\begin{split}
\cH^2(Q,P)=\sum_{n_1+n_2={\rm odd}} A_{n_1n_2} Q^{n_1}P^{n_2}, \ \ \ \
A_{n_1n_2}=A_{n_2n_1},
\end{split}
\label{eq:scheme1}
\]
with numerical coefficients $A_{n_1n_2}$.
Then, if $P=-Q$, one obtains
\[
\begin{split}
\left. \cH^2(Q,P)\right |_{P=-Q}&=\left.\frac{1}{2} \left( \cH^2(Q,P)+\cH^2(P,Q)\right)\right |_{P=-Q}\\
&=\frac{1}{2} \left( \cH^2(Q,-Q)+\cH^2(-Q,Q)\right) \\
&=\frac{1}{2} \sum_{n_1+n_2={\rm odd}} A_{n_1n_2} \left( (-1)^{n_1} +(-1)^{n_2}\right)
Q^{n_1}Q^{n_2} \\
&=0,
\end{split}
\]
because $(-1)^{n_1} $ and $(-1)^{n_2}$ take opposite signs for $n_1+n_2={\rm odd}$.
It is obvious that one can do the same discussion for the exact expression instead of the schematic one above.
Therefore, the Hamiltonian constraint is satisfied for $P=-Q$.

Let us next consider the secondary constraint obtained from the Poisson bracket of two $\cH^2$'s:
\[
\cH^{sec}_{a_1a_2}(Q,P)=\{ \cH^2_{a_1}(Q,P),\cH^2_{a_2}(Q,P) \}.
\]
From \eq{eq:symh2} and the anti-symmetric property of the Poisson bracket, one can easily find that 
\[
\cH^{sec}_{a_1a_2}(Q,P)=-\cH^{sec}_{a_1a_2}(P,Q).
\]
One can also find
\[
\hbox{Order}(\cH^{sec}_{a_1a_2})=\hbox{even}.
\]
Therefore, schematically, 
\[
\begin{split}
\cH^{sec}(Q,P)=\sum_{n_1+n_2={\rm even}} A^{sec}_{n_1n_2} Q^{n_1}P^{n_2}, \ \ \ \
A^{sec}_{n_1n_2}=-A^{sec}_{n_2n_1} .
\end{split}
\label{eq:scheme2}
\]
Then, one can prove that 
\[
\begin{split}
\left. \cH^{sec}(Q,P)\right|_{P=-Q}&=\left. \frac{1}{2} \left( \cH^{sec}(Q,P)-\cH^{sec}(P,Q)\right)\right|_{P=-Q} \\
&=\frac{1}{2} \left( \cH^{sec}(Q,-Q)-\cH^{sec}(-Q,Q)\right) \\
&=\frac{1}{2} \sum_{n_1+n_2={\rm even}} A^{sec}_{n_1n_2} \left( (-1)^{n_1} -(-1)^{n_2}\right)
Q^{n_1}Q^{n_2} \\
&=0.
\end{split}
\]
Therefore, $P=-Q$ satisfies the constraint $\cH^{sec}_{a_1a_2}=0$.

This can be continued for any order of $\cH^2$. Let us define the secondary constraint at order $m$ as 
\[
\cH^{sec}_{a_1a_2\cdots a_m}=\{\cH^2_{a_1} 
\{\cH^2_{a_2}\{\cdots,\cH^2_{a_m}\}\cdots \}.
\label{eq:seccon}
\]
Then, it is easy to show that they have the schematic form 
\eq{eq:scheme1} for odd $m$, and \eq{eq:scheme2} for even $m$.
By applying the same argument as above,  all the secondary constraints can be shown to 
be satisfied by $P=-Q$.

The other constraint ${\cal J}_{ab}=0$ is obviously satisfied by $P=-Q$. 
It is also obvious that all the secondary constraints 
generated from the Poisson brackets among ${\cal J}_{ab}$ and ${\cal H}_a,{\cal H}^{sec}_{a_1\dots a_m}$ 
are satisfied, because ${\cal J}_{ab}$ are the $SO(N)$ generators and just rotate them.
 
\subsection{Numerically checked solutions with general $\beta_i$ for $N=2$}
The solutions for general $N$ in the previous subsections have a restriction on the possible values of $\beta_i$. 
In this subsection, we will present another kind of solutions 
which are valid for any values of $\beta_i$, but only valid for $N=2$.

The solution is constructed by embedding the $N=1$ solution as follows:
\[
Q_{111}=Q,\ P_{111}=P, \ \hbox{the other components}=0,
\label{eq:QPone}
\]
where $Q,P$ have the relation \eq{eq:solP}.   
The constraint ${\cal J}_{ab}=0$ is trivially satisfied for general $N$. And, if ${\cal H}_a={\cal H}^{sec}_{a_1\cdots a_m}=0$ 
are all satisfied, 
all the secondary constraints generated from the Poisson bracket with ${\cal J}_{ab}$ are also satisfied, because ${\cal J}_{ab}$
just rotate them. 
Our numerical observation is that, for $N=2$, \eq{eq:QPone} gives a solution to all the constraints. 
This has been checked by the following numerical evaluations for various values of $\beta_i$:
\begin{itemize}
\item[(i)]
Checking ${\cal H}^2_a=0$ and ${\cal H}^{sec}_{a_1a_2\cdots a_m}=0$ 
up to $m \leq 7$.
\item[(ii)]
Numerically solving the equation of motion of $Q,P$, 
and checking whether ${\cal H}^2_{a}(Q(t),P(t))=0$ remains true over the range of $t$ considered.
\end{itemize} 

For example, we have considered $\beta_1=1,\beta_2=2,\beta_3=-3,\beta_4=-4$. 
In this case, we can take $Q=-\sqrt{\frac{3}{2}},P=2$ satisfying \eq{eq:solP}.
As for (i), we have numerically checked whether the secondary 
constraints are satisfied by \eq{eq:QPone} up to $m\leq 7$. 
Since it would take too long to compute all the components of the secondary constraints, 
we consider the secondary constraints contracted with randomly generated vectors $R^i$:
\[
{\cal H}_{m}=
R^1_{a_1} R^2_{a_2}\cdots R^m_{a_m} {\cal H}^{sec}_{a_1a_2\cdots a_m},
\]  
where $R^i$ have components of randomly generated real numbers between -1 and 1.
This random checking would be enough for the consistency check.
The results are $|{\cal H}_2|,|{\cal H}_{3}|\sim 0, |{\cal H}_{4}|\sim 10^{-13}, 
|{\cal H}_{5}|\sim 10^{-11}, |{\cal H}_{6}|\sim 10^{-10}, 
|{\cal H}_{7}|\sim 10^{-9}$,
which are numerically consistent with zero. 

\begin{figure}
\begin{center}
\includegraphics[scale=.4]{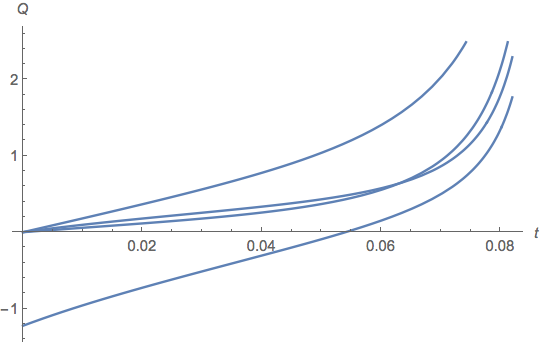}
\hfil
\includegraphics[scale=.4]{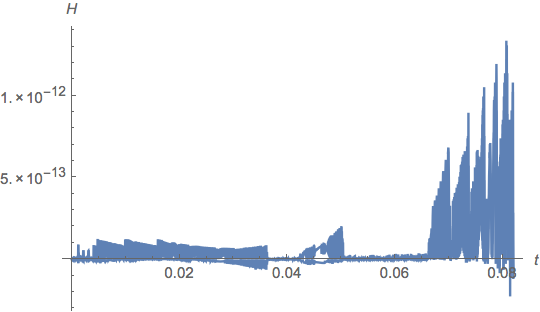}
\caption{The left figure shows the time-dependence of the four independent components of $Q(t)$.
The lowest line is $Q_{111}$ and the others are $Q_{112},Q_{122},Q_{222}$. 
They seem to diverge at $t^*\sim 0.082$.
The right figure shows the time-dependence of ${\cal H}^2_{1},{\cal H}^2_{2}$. 
They are consistent with zero throughout the solved region of $t$.}
\label{fig:plotN=2}
\end{center}
\end{figure}

As for the check (ii), we have numerically solved the equation of motion,
\[
\begin{split}
\frac{d}{dt}Q_{abc}=\{ Q_{abc}, n_d {\cal H}^2_d\}, 
\hspace{10mm}
\frac{d}{dt} P_{abc}=\{ P_{abc}, n_d {\cal H}^2_d\} .
\end{split}
\]
with $n_a=-1$, where the initial values are taken to be \eq{eq:QPone} with the same values of the parameters. 
In Figure~\ref{fig:plotN=2}, we plot the solution $Q(t)$ and ${\cal H}^2(Q(t),P(t))$. 
The solution has a non-trivial time-dependence and seems to diverge at $t^* \sim 0.082$. 
Throughout the solved region of $t$, the Hamiltonian is kept suppressed 
within the order of $10^{-12}$, which is numerically consistent with zero.  

We have also checked whether \eq{eq:QPone} is a solution in $N=3$ or not.
The answer is negative. We have considered the same values of 
the parameters, as in the $N=2$ case.  We have obtained $|{\cal H}_{6}|\sim 10^4$,
while $|{\cal H}_2|\sim0,|{\cal H}_{3}|, |{\cal H}_{4}|\sim 10^{-13}, 
|{\cal H}_{5}|\sim 10^{-11}$. Though it is interesting to find that the 
constraints are satisfied up to the fifth order, it is largely violated at the 
sixth order.

\section{Relation between mCTM and tensor network dynamics}
\label{sec:network}
In the Hamilton formalism of general relativity, spacetime is described by a 
dynamical evolution of a space. Then, when a space is described in terms of randomly 
connected tensor networks (RCTN) as in \cite{Chen:2016xjx}, 
spacetime should be regarded as a consequence of a dynamical evolution of 
RCTN\footnote{In \cite{May:2016dgv}, dynamical evolutions of tensor network spaces
are discussed from a different perspective.}.  
In fact, the Hamiltonian of CTM can be interpreted as an operator to generate networks.
This is used to argue that the Hamiltonian of CTM generates a sort 
of renormalization group flow of RCTN \cite{Sasakura:2014zwa,Sasakura:2015xxa}, 
and it would be the background reason for the relation between a general relativistic system and 
CTM \cite{Sasakura:2015pxa,Chen:2016ate}.

\begin{figure}
\begin{center}
\includegraphics[scale=.5]{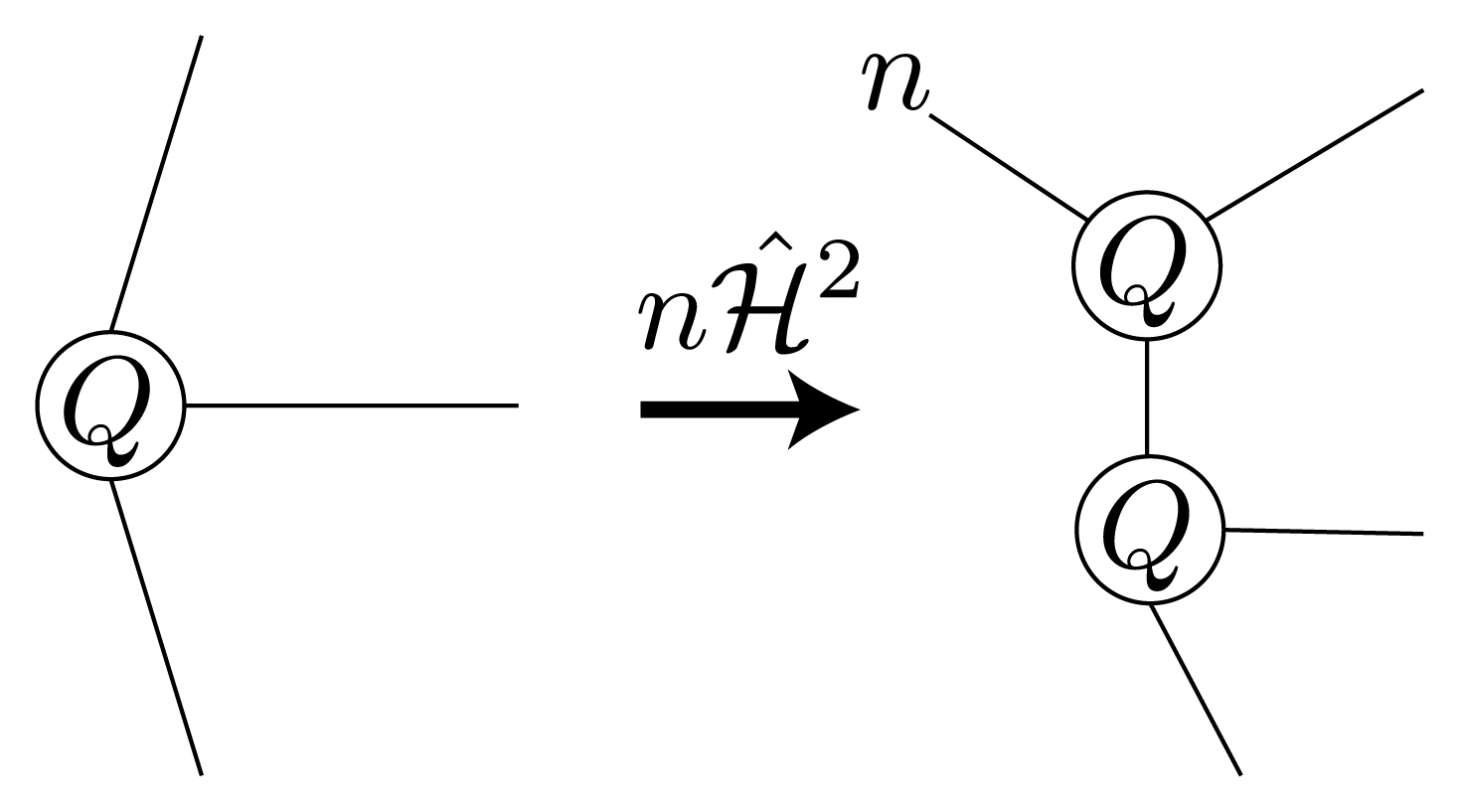}
\hspace{1cm}
\includegraphics[scale=.5]{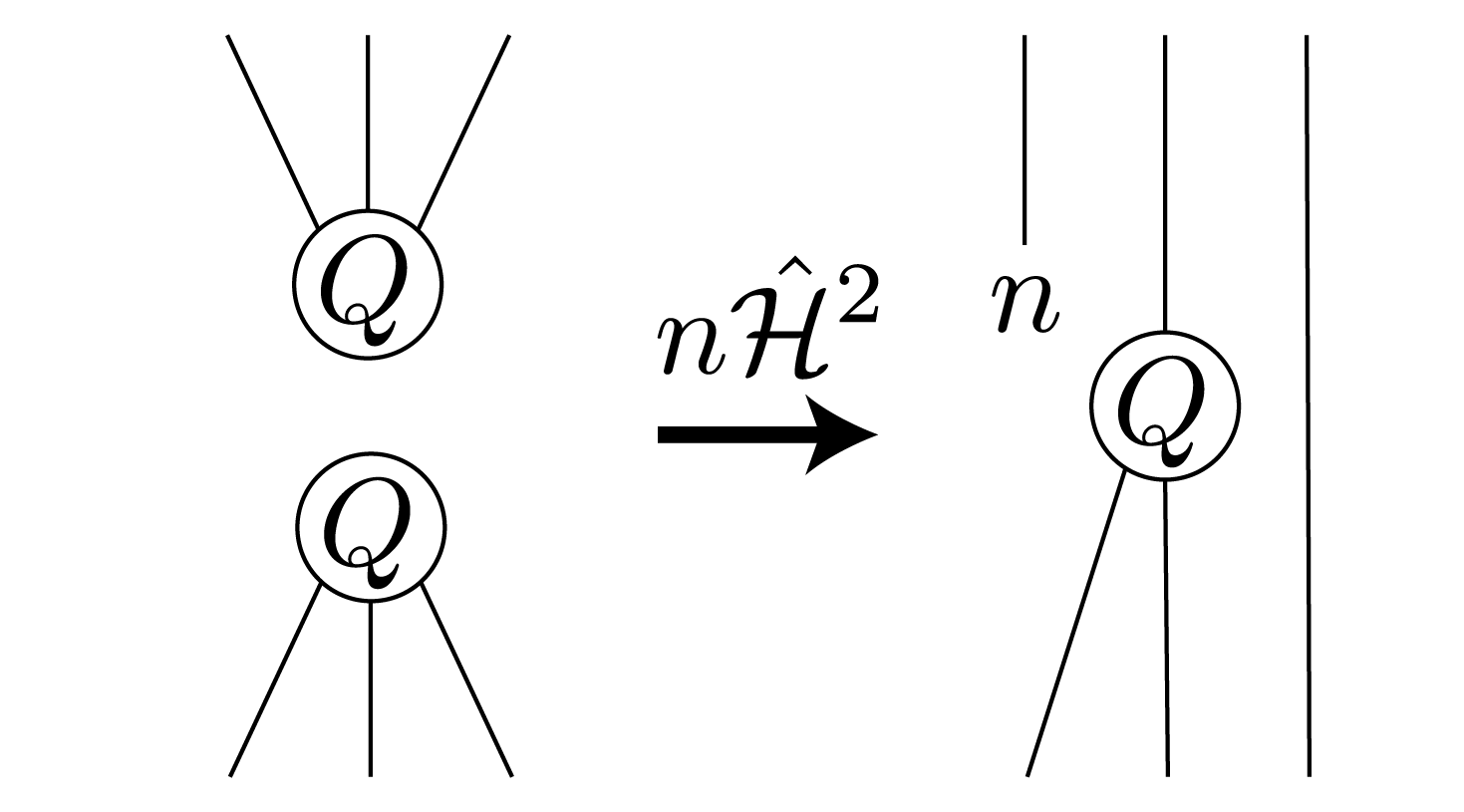}
\end{center}
\caption{In the left, the operation \eq{eq:secondop} is represented in a graphical manner. 
It inserts a tensor on a connection. In the right, the operation of the other cubic term in mCTM is represented. This 
merges two tensor vertices into one. }
\label{fig:ham2op}
\end{figure}

The relation between mCTM and the network dynamics can 
be studied by considering the operations implied by the Hamiltonian ${\cal H}^2$ in \eq{eq:ham2}. 
Let us first consider the second term of ${\cal H}^2$. 
Taking the Poisson bracket between $Q_{abc}$ and the second term, one obtains
\[
\{Q_{abc}, n_{d} Q_{def}Q_{egh}P_{fgh} \}=\sum_{\sigma} n_d Q_{d e \sigma_a}Q_{e \sigma_b \sigma_c}.
\label{eq:secondop}
\]
This operation can be graphically described by the left figure of Figure~\ref{fig:ham2op}.
Therefore, if we apply this operation to a tensor network of $Q$, 
it inserts one $Q$ randomly on one of its links. This changes the size 
of the tensor network by one tensor. In fact, this operation is used to 
change the size $n$ of the network in the partition function of RCTN \eq{eq:partRCTN},
and we argued that the Hamiltonian of CTM generates an RG-like flow of RCTN in the infrared direction 
\cite{Sasakura:2014zwa,Sasakura:2015xxa}.

What is new in the mCTM is the coexistence of the first and second 
terms of ${\cal H}^2$, while there exists only one of them in the case of CTM
(depending on the roles of $Q,P$). 
To describe the operation, it is more convenient to move to the 
quantum framework from the classical one. The first term of $\hat {\cal H}^2$ 
in \eq{eq:qham2} can be represented by $Q_{cde}D_{abc}D_{bde}$
in the $Q$-representation (Here, the normal ordering term is absorbed into $\beta_3$.). 
Then, we see that this gives an operation which is shown in the right figure of Figure~\ref{fig:ham2op}:
it merges two $Q$'s into one. It is obvious that similar things can be 
discussed for $\hat {\cal H}^3$ in \eq{eq:qham3} by applying it on the Fock states.

\begin{figure}
\begin{center}
\includegraphics[scale=.5]{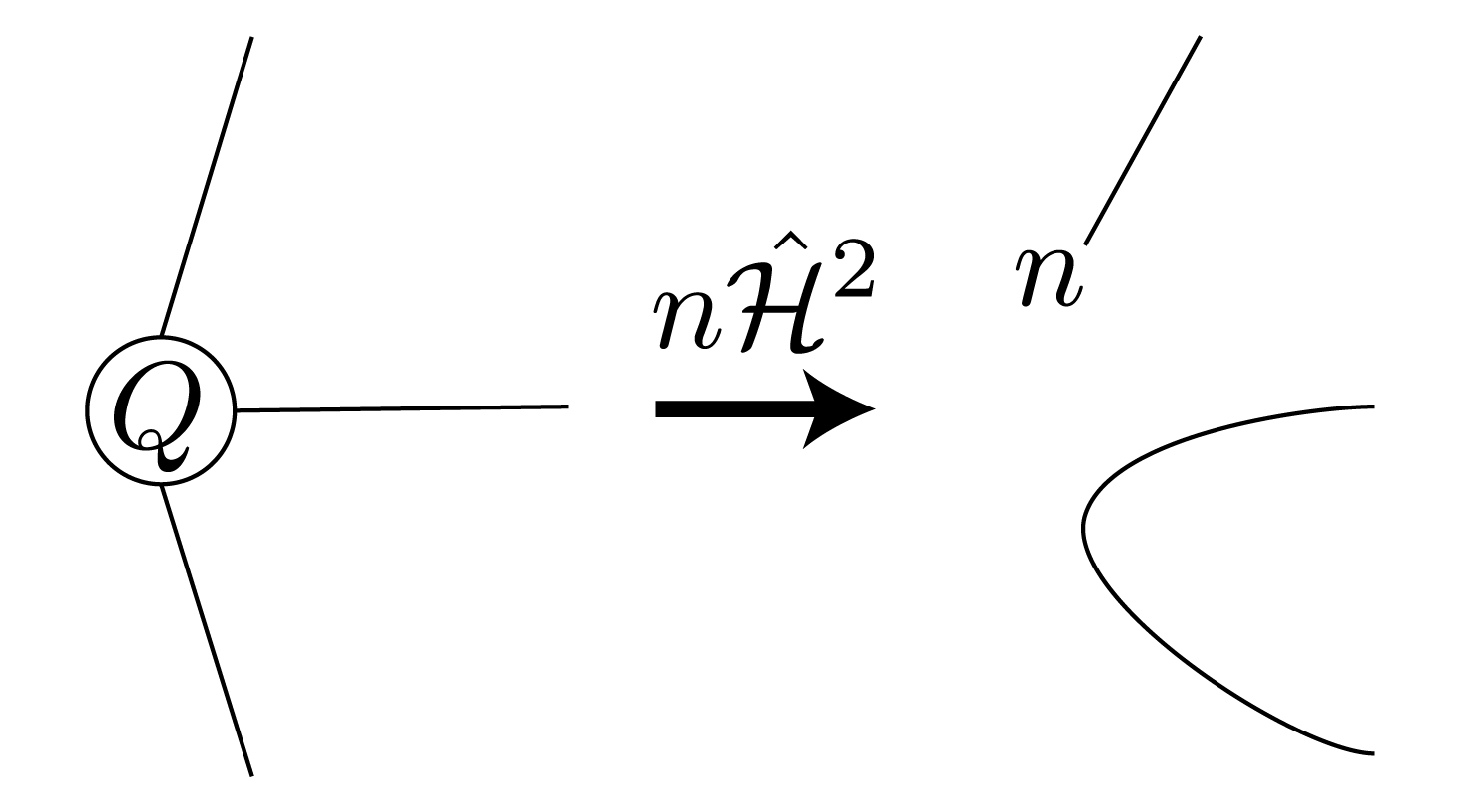}
\hspace{1cm}
\includegraphics[scale=.5]{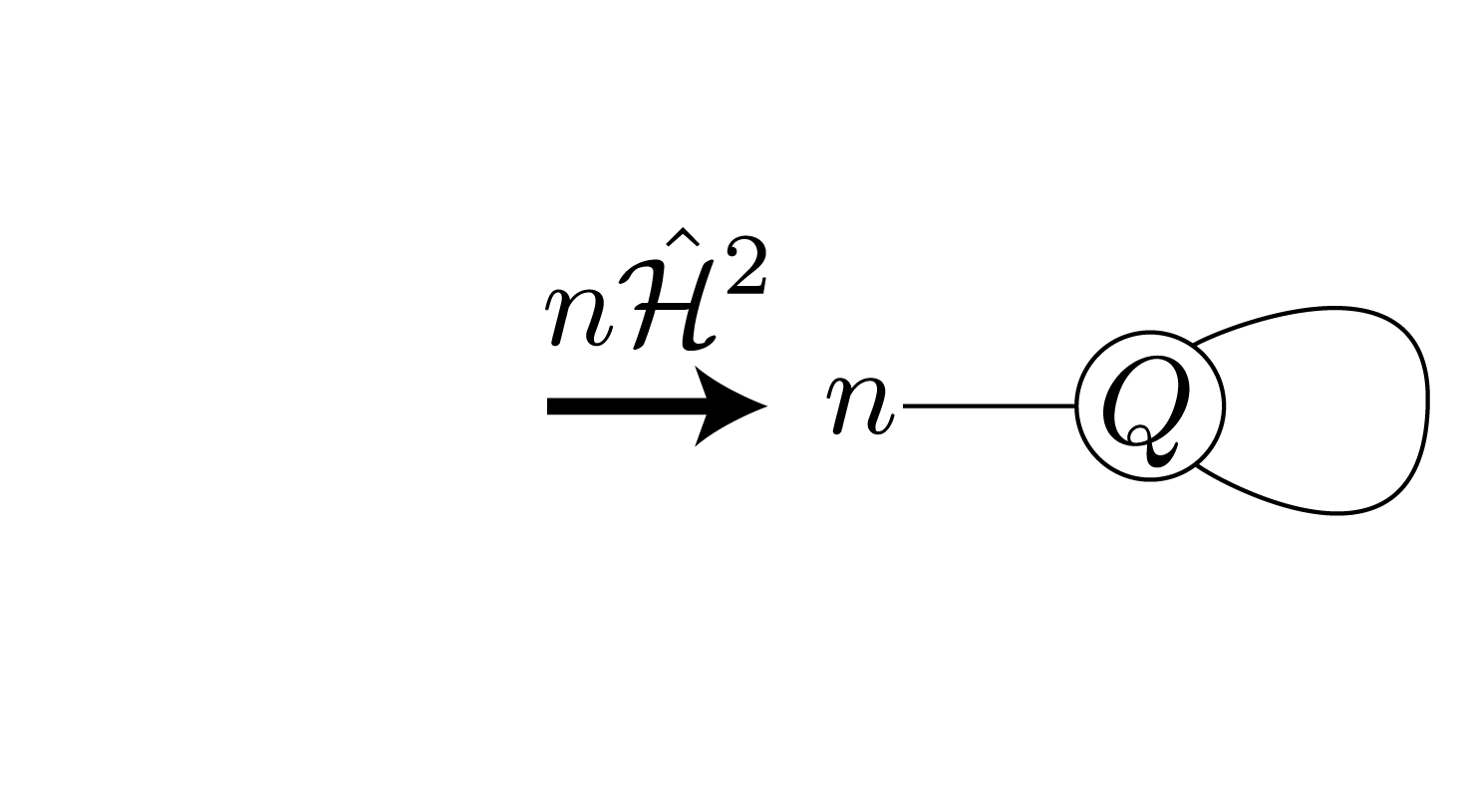}
\end{center}
\caption{The operations corresponding to the third and the fourth terms of $\hat {\cal H}^2$.}
\label{fig:hatham2op}
\end{figure}

In Figure~\ref{fig:hatham2op}, the operations corresponding to the third 
and fourth terms of $\hat{\cal H}^2$ (and $\hat{\cal H}^3$) are also shown.

\section{Summary and outlook}
\label{sec:summary}
In our previous paper \cite{Chen:2016ate}, we have analyzed a formal 
continuum limit of the classical equation of motion of the canonical tensor 
model (CTM) up to the fourth-order terms of spatial derivatives in a derivative expansion of the tensor, 
and have found that it is equivalent to the classical equation of motion 
of a coupled system of general relativity and a scalar 
field in the Hamilton-Jacobi formalism. This achievement suggests the 
existence of a ``mother" theory of CTM (mCTM) which derives CTM through the Hamilton-Jacobi
procedure. Such an mCTM would be more directly related to the gravitational 
system (see Figure~\ref{fig:scheme}), without resort to the Hamilton-Jacobi procedure.
In this paper, we have successfully found an mCTM, and have performed some initial studies
of its properties. The quantization is also straightforward, as in the case of CTM. 

The clearest characterization of the Hamiltonian of mCTM we have found
is that it contains both the creation and annihilation 
operations of tensor vertices on tensor networks, as described in Section~\ref{sec:network}.
This is in sharp contrast to CTM, whose Hamiltonian contains only one of these operations 
(the two choices depend on which of $Q$ or $P$ represents tensor networks.). 
Therefore, while CTM shows the characteristics of first-order differential equations,
such as what appear in the renormalization group flow of randomly connected tensor 
networks \cite{Sasakura:2014zwa,Sasakura:2015xxa} and 
the Hamilton-Jacobi formalism of a gravitational system \cite{Chen:2016ate}, 
mCTM can be expected to show more characteristics of dynamical systems
governed by second-order differential equations.
In fact, the oscillatory behavior of a physical wave function in Figure~\ref{fig:caseii} can be 
thought as an appearance of such characteristics, compared with the
rather monotonous physical wave functions of CTM \cite{Narain:2014cya,Sasakura:2013wza}.
The characteristics would also make us expect that the continuum limit of mCTM can be directly related to 
a gravitational system, without resort to the Hamilton-Jacobi formalism. 
This is certainly an important subject in future study.

A serious disadvantage of mCTM we have found is that we do not know the 
full structure of the constraints and their algebra. Nonetheless, in this paper,
we have succeeded in finding some exact physical wave functions and classical phase spaces
which solve the primary and all the secondary constraints without knowing the precise expressions of the latter.
This is enough to show that mCTM is not vacuous and is a physically interesting system.
On the other hand, it is obvious that the lack of such knowledge is a 
serious obstacle to further study of mCTM, and probably, the most important 
problem is that we do not understand the gauge symmetry of mCTM,
which should be represented in the constraints and their algebra. 
The gauge symmetry should provide a tensor model correspondence 
to the covariance in general relativity, and would be essentially important 
in formulating quantum gravity based on this principle. 
This would also be an important future subject.

Another direction that is worth exploring is considering an 
$OSp$-extension of mCTM. In \cite{Narain:2015owa} we considered 
a super-extension of ordinary CTM by incorporating fermionic 
degrees of freedom. This was a straightforward generalization 
of CTM, but we realized that it contains states having negative 
norms, which arose skepticism regarding the $OSp$-extension of CTM. 
It will be interesting to see whether such an extension can also 
be implemented in mCTM, and whether such a generalization might 
resolve the issues of negative norm states. This is an 
interesting direction which will be pursued in future. 

As obtained in this paper, mCTM also shows an analytical solvability 
of the physical wave functions as the case of CTM \cite{Narain:2014cya,Sasakura:2013wza}. 
Presently, this aspect is rather mysterious, and the 
reason behind should be revealed in future study.

\vspace{1cm}
%\section*{Acknowledgements}
%%%%%%%%%%%%%%%%%%%%%%%%%%%%%%%%%%%%%%%%%%%%%%%
\centerline{\bf Acknowledgements} 
The work of N.S. is supported in part by JSPS KAKENHI Grant Number 15K05050. 
N.S. would like to thank the members of KITPC for the hospitality, while he 
stayed there and part of this work was done. 
%%%%%%%%%%%%%%%%%%%%%%%%%%%%%%%%%%%%%%%%%%%%%%%%

\begin{appendix}
\section{$SL(2,R)$ transformation of ${\cal H}^1$ to ${\cal H}^2$}
\label{app:trans1to2}
By putting \eq{eq:sl2r} to \eq{eq:ham1}, one obtains
\[
\begin{split}
{\cal H}^1{}'=&
(\alpha_2z_2^3 + \alpha_1 z_2 z_4^2)P_{abc}P_{bde}P_{cde}
+(2 \alpha_2 z_1 z_2^2 + 2 \alpha_1 z_2 z_3 z_4)P_{abc}P_{bde}Q_{cde} \\
&+(\alpha_2 z_1^2 z_2 + \alpha_1 z_2 z_3^2)P_{abc}Q_{bde}Q_{cde}
+(\alpha_2 z_1 z_2^2 + \alpha_1 z_1 z_4^2)Q_{abc}P_{bde}P_{cde}\\
&+(2 \alpha_2 z_1^2 z_2 + 2 \alpha_1 z_1 z_3 z_4)Q_{abc}Q_{bde}P_{cde}
+(\alpha_2 z_1^3 + \alpha_1 z_1 z_3^2)Q_{abc}Q_{bde}Q_{cde}\\
&+\left(\alpha_3 z_3+\alpha_4 z_1 \right) Q_{abb}+\left(\alpha_3 z_4 +\alpha_4 z_2\right) P_{abb}.
\end{split}
\label{eq:hprime}
\]
The form of ${\cal H}^2$ in  \eq{eq:ham2} can be obtained by imposing\footnote{The other possibility $z_1=z_2=0$ contradicts 
the determinant condition.}
\[
\alpha_2z_2^2 + \alpha_1 z_4^2=0,\ \ \ 
\alpha_2 z_1^2 + \alpha_1 z_3^2=0.
\]
A real solution requires the relative sign between $\alpha_1$ and $\alpha_2$ to be minus.
Then, from the condition that the determinant is 1 in \eq{eq:sl2r}, the solution is obtained as 
\[
z_2=\pm \frac{1}{2 z_1} \sqrt{-\frac{\alpha_1}{\alpha_2}},\ z_3=\mp z_1 \sqrt{-\frac{\alpha_2}{\alpha_1}},\ 
z_4=\frac{1}{2z_1}
\label{eq:solz}
\] 
with free choice of $z_1$. By putting the solution \eq{eq:solz} to \eq{eq:hprime}, we obtain
\[
\begin{split}
{\cal H}_a^2=&-\frac{\alpha_1}{z_1} P_{abc}P_{bde}Q_{cde}\mp 2 z_1 \hbox{Sgn}(\alpha_1)\sqrt{-\alpha_1 \alpha_2} Q_{abc}Q_{bde}P_{cde}\\
& +z_1 \left( \alpha_4 \mp \alpha_3 \sqrt{-\frac{\alpha_2}{\alpha_1}}\right)  Q_{abb}
+\frac{1}{2 z_1} \left(\alpha_3 \pm \alpha_4 \sqrt{-\frac{\alpha_1}{\alpha_2}}\right)P_{abb}.
\end{split}
\]

\section{$N=1$ wave function}
\label{app:n=1wavefn}
The confluent hypergeometric function $F(A_1,A_2;x)$ can be defined by the following perturbative expansion around $x=0$,
\[
F(A_1,A_2;x)=\sum_{n=0}^\infty \frac{(A_1)_n}{(A_2)_n}\frac{x^n}{n!},
\label{eq:expF}
\]
where 
\[
(A)_n=A (A+1)(A+2)\cdots (A+n-1)=\frac{\Gamma(A+n)}{\Gamma(A)}.
\label{eq:defAn}
\]
The asymptotic formula of the confluent geometric function is given by
\[
\begin{split}
F(A_1,A_2;x)\sim &\frac{\Gamma(A_2)}{\Gamma(A_2-A_1)} (-x)^{-A_1} 
\sum_{n=0}^\infty (-1)^n\frac{(A_1)_n(A_1-A_2+1)_n}{n!\ x^n} \\
&+\frac{\Gamma(A_2)}{\Gamma(A_1)} \, e^x x^{A_1-A_2} \sum_{n=0}^\infty (-1)^n\frac{(1-A_1)_n(A_2-A_1)_n}{n!\ x^n}.
\end{split}
\label{eq:asymF}
\]
From \eq{eq:imbeta} with $N=1$ and \eq{eq:valuesA}, the real parts of $A_1,A_2$ are given by
\[
A_1&=3+\hbox{imaginary},\\
A_2&=-\frac{3}{2}+\hbox{imaginary}.
\]
Therefore, the second series of the expansion in \eq{eq:asymF} is asymptotically diverging 
(note that $e^x$ is just an oscillatory factor in the present case with imaginary $x$), 
while the first term decays in $\sim q^{-6}$.  From \eq{eq:asymF}, 
the condition for the cancellation of the second series in the solution \eq{eq:gensol} is given by \eq{eq:cancelsecond}.

\end{appendix}

%%%%%%%%%%%%%%%%%%%%%%%%%%% 

\end{document}